\documentclass[preprint,aps,showkeys,showpacs]{revtex4}

\usepackage{graphicx}
\usepackage{color}
\input psfig.sty

\newcommand{\bq}{\begin{equation}}
\newcommand{\eq}{\end{equation}}
\newcommand{\bqn}{\begin{eqnarray}}
\newcommand{\eqn}{\end{eqnarray}}
\newcommand{\nb}{\nonumber}
\newcommand{\lb}{\label}

\begin{document}

\title{Gravitational Collapse of an 
Imperfect Non Adiabatic Fluid}

\author{R. Chan}
\email{chan@on.br}
\affiliation{Coordena\c c\~ao de Astronomia e Astrof\'{\i}sica,
Observat\'orio Nacional, Rua General Jos\'e Cristino 77, S\~ao
Crist\'ov\~ao, CEP 20921--400, Rio de Janeiro, RJ, Brazil}

\author{M. F. A. da Silva}
\email{mfasnic@gmail.com}
\affiliation{Departamento de F\' {\i}sica Te\' orica,
Universidade do Estado do Rio de Janeiro,
Rua S\~ ao Francisco Xavier $524$, Maracan\~ a,
CEP 20550--013, Rio de Janeiro, RJ, Brazil} 

\author{C. F. C. Brandt}
\email{fredcharret@yahoo.com.br}  
\affiliation{Departamento de F\' {\i}sica Te\' orica,
Universidade do Estado do Rio de Janeiro,
Rua S\~ ao Francisco Xavier $524$, Maracan\~ a,
CEP 20550--013, Rio de Janeiro, RJ, Brazil} 

\date{\today}

\begin{abstract}
We study the evolution of an
anisotropic shear-free fluid with heat flux and kinematic self-similarity of the second
kind.  We found a class of solution to the Einstein field equations by
assuming that the part of the tangential pressure which is explicitly time
dependent of the fluid is zero
and that the fluid moves along time-like geodesics. 
The energy conditions, geometrical and physical properties
of the solutions are studied. 
The energy conditions are all satisfied at the beginning of the collapse
but when the system approaches the singularity the energy conditions are 
violated, allowing for the appearance of an attractive phantom energy.
We have found that, depending on the self-similar parameter 
$\alpha$ and the geometrical radius, they may represent a 
naked singularity. 
We speculate that the apparent horizon disappears due to the emergence of 
exotic energy  at the end of the collapse, or due to the characteristics of 
null acceleration systems as shown by recent work.
\end{abstract}  

\keywords{gravitational collapse, self-similarity, heat flux, anisotropic fluid,
black hole, naked singularity}

\pacs{04.20.Jb,04.20.Dw,04.40.Dg,04.70.Bw,04.25.dc}

\maketitle

\section{Introduction}

There are two exciting subjects in General Relativity (GR) nowadays.
One of them is related to the Cosmic Censorship
Conjecture and the other one is Critical Phenomena in the context of the 
Gravitational Collapse.

One of the most important problems in gravitation theory is the final
state
of a collapsing massive star, after it has exhausted its nuclear fuel.
Despite
many efforts over the last four decades, our understanding is still
limited to several conjectures, such as, the cosmic censorship conjecture
\cite{Penrose}, and the hoop conjecture \cite{Thorne}. To the former,   many
counter-examples have been found \cite{Joshi}-\cite{Joshi3}, although  
it is still not clear
whether  those particular examples are stable and generic. To the latter,
no counter-examples have been found so far in four-dimensional Einstein's
Theory of gravity, although it has been shown
that this is no longer the case in five dimensions \cite{NM01,NM011}.
On the other hand, Choptuik's discovery on critical phenomena in gravitational
collapse near the threshold of black holes formation gave us deep
insight to the non-linearity of the Einstein field equations
\cite{Choptuik93}. Now critical phenomena in gravitational
collapse already become a well-established subarea in GR
\cite{Gun00,Wang01}.

Many works have been done so far in this area, it can be seen
that  the critical solutions, which separate the collapse that
form black holes from the one that does not form black holes, can
have discrete self-similarity (DSS), continuous self-similarity
(CSS), or none of them, depending on both matter fields and
regions of initial data space. The collapse can be type II if the
black holes start to form with zero mass.  In this case it is
found that the critical solutions have either   DSS or CSS. The
collapse can also be type I, that is,   the formation of black holes
starts with a finite non-zero mass. It is found that in the
latter case the critical solutions have no self-similarities,
neither DSS nor CSS \cite{Gun00,Wang01}.
It is interesting that even the collapse is not critical,
but if it has self-similarity, it is found that the formation of
black holes can still start with zero-mass
\cite{Jaimepaper,Jaimepaper1,Jaimepaper2,Jaimepaper3,Jaimepaper4}.
For details, we would like to refer readers to
\cite{Jaimepaper,Jaimepaper1,Jaimepaper2,Jaimepaper3,Jaimepaper4,CC97,CC971,CC972,CC973,Gundlach}
and references therein.

We and other authors have studied gravitational collapse of 
anisotropic fluids
with kinematic self-similarities in four-dimensional spacetimes
\cite{Carlos02}\cite{Brandt}\cite{artigonossofluxocalor2003}
\cite{CdSVR}
and references therein.
For example, Brandt et al. (2003) \cite{artigonossofluxocalor2003} have 
analyzed the collapse of an anisotropic 
fluid with self similarity of the first kind, with no heat flux.  They have 
shown that the system formed a black holes at the end.
Besides, Brandt et al. (2006) \cite{Brandt} have studied the collapse of the 
second 
type for values $\alpha = 1$ and $3/2$, with equation of state with radial 
pressure proportional to the energy density, tangential pressure is zero and
no heat flux.
They have shown that there is formation of black hole ($\alpha = 3/2$) and 
naked singularity ($\alpha = 1$).
Another naked singularity
appears at the end of the collapse for an anisotropic fluid with heat flow
and with self-similarity of the second kind \cite{CdSVR}.

In this work, we have studied general solutions of the Einstein's equations
for a second kind self-similar anisotropic shear-free fluid with heat flux.  
We have analyzed some 
particular cases, which gives us completely different final states, including a
naked singularity
, representing a new counter-example to the cosmic censorship. 
The paper is organized as follows.
In Section 2 we present the Einstein field equations. In Section 3 we
present a class of exact solutions that represents an anisotropic
fluid moving along time-like geodesics.  The ingoing, outgoing null congruence scalar
expansions and the energy conditions are analyzed \cite{HE73}. 
The energy conditions are all satisfied at the beginning of the collapse
but when the system approaches the singularity the energy conditions are 
violated, allowing for the appearance of an attractive phantom energy 
\cite{Chan2009}.
Finally, in Section 4 we present the conclusions.

\section{The Field Equations}

The general metric of spacetimes with spherical symmetry  can 
be cast in the form,
\bq
\lb{genmetric}
ds^2 = r_1^2 \left[e^{2 \Phi(t,r)}dt^2 - e^{2 \Psi(t,r)} dr^2
- r^2 S^2(t,r) d\Omega^2\right],
\eq
where $d\Omega^2 =  d\theta^2 + \sin^2\theta d\phi^2 $, and
$r_{1}$ is a constant with the dimension of length. Then, we can
see that the coordinates $t, r, \theta$ and $\phi$, as well as the
functions $\Phi, \Psi$ and $S$ are all dimensionless.

Self-similar solutions of the second kind  are given by
\bq
\lb{seckind}
\Phi(t,r) = \Phi(x),\;\;\; \Psi(t,r) = \Psi(x),\;\;\;
S(t,r) = S(x),
\eq
where
\bq
\lb{ss}
x \equiv \ln \left[ \frac{r}{(-t)^{1/\alpha}} \right],
\eq
and $\alpha$ is a dimensionless constant.  
The  general energy-momentum tensor of an anisotropic fluid
can be cast in the form
\bq
\lb{emtgen}
T_{\mu \nu} = \rho u_{\mu} u_{\nu}+
p_{t} (\theta_{\mu} \theta_{\nu} + \phi_{\mu}\phi_{\nu})
+ p_{r} n_{\mu}  n_{\nu} + q( u_{\mu} n_{\nu} + u_{\nu} n_{\mu}),
\eq
where $u^{\mu}$ denotes the   four-velocity of the fluid and $n^a$
is a unit spacelike vector orthogonal to $u^a$, while
$\theta_{\nu}$ and $\phi_{\mu}$ denote the unit vectors in the
tangential directions.  Then, we can see that $\rho $ is the
energy density of the fluid measured by observers comoving with
the fluid, $p_t$ and $p_r$ are respectively the tangential and
radial pressures and $q$ is the radial heat flux.  
In the comoving coordinates, we have
\bqn
u_{\mu} &=& e^{\Phi(x)} \delta_{\mu}^t,  \;\;\;\;
n_{\mu} = e^{\Psi(x)} \delta^r_{\mu},     \nb\\
\theta_{\mu} &=& r S(x)\delta_{\mu}^{\theta},  \;\;\;\;
\phi_{\mu}= r S(x) \sin\theta \delta_{\mu}^\phi .
\eqn

Defining 
\bq 
y \equiv \frac{\dot S}{S}, 
\label{ySS}
\eq 
where the symbol dot over the variable denotes 
differentiation with respect to $x$.
We find that the non-null  components of the Einstein tensor in
the coordinates $\{t, r, \theta, \phi\}$ can be written as 
\bqn
\lb{C.4a}
G_{tt} &=& - \frac{1}{r^{2}}e^{2(\Phi -\Psi)} \left[2\dot y + y(3y +
4) + 1 - 2(1+y)\dot \Psi - S^{-2}e^{2\Psi}\right]\nb\\
& & + \frac{1}{\alpha^2 t^{2}}(2\dot \Psi + y)y,\\
\lb{C.4b}
G_{tr} &=&  \frac{2}{\alpha t r}
\left[\dot y + (1+y)(y - \dot \Psi) - y\dot \Phi\right],\\
\lb{C.4c}
G_{rr} &=& \frac{1}{r^{2}}\left[
2(1+y)\dot \Phi + (1 + y)^{2}  - S^{-2}e^{2\Psi}\right]\nb\\
& & - \frac{1}{\alpha^2 t^{2}}e^{2(\Psi - \Phi)}
\left[2\dot y + y\left(3y - 2\dot \Phi + 2\alpha\right)\right],\\
\lb{C.4d}
G_{\theta\theta} &=&  S^{2}e^{-2\Psi}\left[\ddot \Phi + \dot y +
\dot \Phi\left(\dot \Phi - \dot \Psi + y\right) +
\left(1 + y\right)\left(y - \dot \Psi\right)\right]\nb\\
& &  - \frac{r^{2}S^{2}}{\alpha^2 t^{2}}e^{-2\Phi}\left[\ddot \Psi + \dot y
+ y^{2} - \left(\dot \Psi + y\right)\left(\dot \Phi - \dot \Psi -
\alpha \right)\right], \\
G_{\phi\phi} &=&  G_{\theta\theta} \sin^2 \theta
\eqn
and the components of the energy-momentum tensor are
\bqn
T_{tt} &=& \rho e^{2\Phi}, \\
T_{tr} &=& q e^{\Phi+\Psi}, \\
T_{rr} &=& p_r e^{2\Psi}, \\
T_{\theta\theta} &=& p_t r^2 S^2, \\
T_{\phi\phi} &=& p_t r^2 S^2 \sin^2 \theta, 
\eqn
where in writing the above expressions we have set $r_{1} = 1$.
From these expressions  we find that the Einstein field equations
$G_{\mu\nu} = T_{\mu\nu}$ can be written as
\bqn
\lb{physicalquant1}
\rho &=& \frac{V^{(1)}(x)}{r^2} + \frac{V^{(2)}(x)}{t^2},  \\
p_{r} &=& \frac{P^{(1)}_{r}(x)}{r^2} + \frac{P^{(2)}_{r}(x)}{t^2}, \nb \\
p_{t} &=& \frac{P^{(1)}_{t}(x)}{r^2 } + \frac{P^{(2)}_{t}(x)}{t^2 }, \nb \\
q &=& \frac{Q(x)}{tr},\nb
\eqn
where
\bqn
\lb{physicalquant2}
V^{(1)}(x) &=& \frac{1}{r^{2}}e^{-2\Psi} \left[2\dot y + y(3y +
4) + 1 - 2(1+y)\dot \Psi - S^{-2}e^{2\Psi}\right], \nb\\
V^{(2)}(x) &=& \frac{1}{\alpha^2}e^{-2\Phi} y [ y + 2 \dot{\Psi}],  \nb\\
P^{(1)}_{r}(x) &=& -\frac{1}{S^2} + e^{-2\Psi}(1 + y)[ 1 + y + 2 \dot{\Phi}], \\
P^{(2)}_{r}(x) &=& - \frac{1}{\alpha^2}e^{-2\Phi} [ 2 \dot{y} + 2\alpha y + 3 y^2 - 2 y \dot{\Phi}], \nb\\
P^{(1)}_{t}(x) &=& e^{-2\Psi}\left[\ddot \Phi + \dot y + \dot \Phi\left(\dot \Phi - \dot \Psi + y\right) +
\left(1 + y\right)\left(y - \dot \Psi\right)\right],    \nb\\
P^{(2)}_{t}(x) &=& \frac{1}{\alpha^2 }e^{-2\Phi}\left[\ddot \Psi + \dot y 
+ y^{2} - \left(\dot \Psi + y\right)\left(\dot \Phi - \dot \Psi - \alpha \right)\right] \nb \\
Q(x) &=& \frac{2}{\alpha}e^{-(\Phi+\Psi)}\left[ \dot{y} - (1+y)(\dot{\Psi} - y) - y \dot{\Phi} \right]. \nb
\eqn

In the next section we have solved the Einstein's equations.

\section{Geodesic Shear-free Model with an Equation of State}

We study now the solutions of anisotropic
fluid with self-similarity in a geodesic model, that is, a
situation in which the acceleration $\dot{\Phi} = 0$, and in
particular we made $\Phi = 0$.  Thus, we can write equations 
(\ref{physicalquant2}) as
\bqn
\lb{V1}
V^{(1)}(x) &=& \frac{1}{S^2 } - e^{-2\Psi} (1+y)^2, \\
\lb{V2}
V^{(2)}(x) &=& \frac{1}{\alpha^2} y [ y + 2 \dot{\Psi}],  \\
\lb{Pr1}
P^{(1)}_{r}(x) &=& -\frac{1}{S^2} + e^{-2\Psi}(1 + y)^2, \\
\lb{Pr2}
P^{(2)}_{r}(x) &=& - \frac{1}{\alpha^2} [ 2 \dot{y} + 2\alpha y + 3 y^2 ], \\
\lb{Pt1}
P^{(1)}_{t}(x) &=& e^{-2\Psi}\left[\dot y + \left(1 + y\right)\left(y - \dot \Psi\right)\right],  \\
\lb{Pt2}
P^{(2)}_{t}(x) &=&  \frac{1}{\alpha^2 }\left[\ddot \Psi + \dot y
+ y^{2} + \left(\dot \Psi + y\right)\left( \dot \Psi + \alpha \right)\right],\\
Q(x) &=&  \frac{2e^{-\Psi}}{\alpha}
\left[\dot y + (1+y)(y - \dot \Psi) \right].
\lb{Q}
\eqn
Then, in principle, we can have solutions with geometrical, but not
physical self-similarity.

In order to obtain a particular solution of the Einstein's equations, 
let us assume the shear-free condition
\bq
\lb{eqstate}
y = \dot \Psi, 
\label{y}
\eq
and the equation of state
\bq
P^{(2)}_{t}(x) = 0.
\label{Pt20}
\eq
This equation corresponds to the part of the tangential pressure which is 
explicitly time dependent, keeping the self-similar variable $x$ as an
independent variable.

From equation (\ref{Q}) we can see that if we have chosen $y=-1$, we would 
not have heat flux.  This particular case has been studied in a previous paper 
\cite{Brandt}.

Using equation (\ref{y}) and (\ref{Pt20}) and substituting into (\ref{Pt2})
we get
\bq
2\ddot \Psi + \dot \Psi(3\dot \Psi+2\alpha)=0,
\eq
which can be solved giving
\bq
\Psi = \frac{2}{3} \ln | 1 - 3 e^{\alpha(x_0 - x)} | + \Psi_0,
\lb{Psi}
\eq
where $x_0$ is an arbitrary integration constant.

From equations (\ref{y}) and (\ref{ySS}) we obtain that
\bqn
\Psi = \ln \left( \frac{S}{S_0} \right),
\eqn
which furnishes
\bqn
\lb{S}
S=S_0 e^{\Psi},
\eqn
where $S_0$ is another arbitrary integration constant.
Substituting equation (\ref{Psi}) into (\ref{S}) we get
\bq
S=S_0 e^{\Psi_0} \left[1 - 3 e^{\alpha(x_0-x)} \right]^{2/3}.
\eq
We can always assume $S_0=1$ and $\Psi_0=0$ without any loss of
generality.

Thus the metric (\ref{genmetric}) takes the form
\bqn
\lb{metric1}
ds^2 = dt^2 -  \left[1 - 3 e^{\alpha x_0} \frac{(-t)}{r^\alpha} \right]^{4/3}
dr^2 - r^2 \left[1 - 3 e^{\alpha x_0} \frac{(-t)}{r^\alpha} \right]^{4/3} d\Omega^2.
\eqn

The geometric radius is given by
\bqn
\lb{geomradius}
R = r S = r \left[1 + 3 e^{\alpha x_0} t r^{-\alpha} \right]^{2/3}.
\eqn

Before looking for solutions of the Einstein field equations, we must take into account
that for the metric to represent spherical symmetry some physical and geometrical conditions
must be imposed \cite{Fatima0}-\cite{Fatima9}.  We impose the regularity 
condition for the gravitational collapse at the center, i.e.,
$\lim_{r \rightarrow 0} R =0$ at least at the initial time.
Thus, we must have that $\alpha \le 0$. 

Using equation (\ref{geomradius}) and the expression for outgoing and ingoing 
null geodesics \cite{Yasuda,Yasuda1,Yasuda2,Yasuda3}
\bqn
\lb{outgoing}
\theta_l = \frac{f}{R} ( R_{,t} +  e^{-\Psi}R_{,r}   ),
\eqn
and
\bqn
\lb{ingoing}
\theta_n = \frac{g}{R} ( R_{,t} -  e^{-\Psi}R_{,r}),
\eqn
we obtain that
\bqn
\lb{outgoing1}
\theta_l = \frac{f}{r \left[1 + 3 e^{\alpha x_0} t r^{-\alpha} \right]^{2/3}}
\left\{ \frac{2 e^{\alpha x_0} r^{1-\alpha}}{ \left[1 + 3 e^{\alpha x_0} t r^{-\alpha} \right]^{1/3}} + 
1 - \frac{2\alpha e^{\alpha x_0} t r^{-\alpha}}
{1 + 3 e^{\alpha x_0} t r^{-\alpha}} \right\},
\eqn
and
\bqn
\theta_n = \frac{g}{r \left[1 + 3 e^{\alpha x_0} t r^{-\alpha} \right]^{2/3}}
\left\{ \frac{2 e^{\alpha x_0} r^{1-\alpha}}{ \left[1 + 3 e^{\alpha x_0} t r^{-\alpha} \right]^{1/3}} - 
1 + \frac{2\alpha e^{\alpha x_0} t r^{-\alpha}}
{1 + 3 e^{\alpha x_0} t r^{-\alpha} } \right\},
\lb{ingoing1}
\eqn
where $f$ and $g$ are positive functions and the comma means partial 
differentiation. Hereinafter we will analyze the expansions without the
factors $f$ and $g$ since these quantities are always
positive for reasonable physical system \cite{artigonossofluxocalor2003}.

The inequalities that are all common to the energy conditions:
\bq
C_1=\left| {\rho}+{\it p} \right| -2\, \left| {\it q} \right|>0, 
\eq
\bq
C_2={\rho}-{\it p}+2\,{\it p_t}+\Delta>0,
\eq
the weak energy conditions are given by
\bq
C_3={\rho}-{\it p}+\Delta>0,
\eq
the dominant energy conditions are written as
\bq
C_4={\rho}-{\it p}>0
\eq
\bq
C_5={\rho}-{\it p}-2\,{\it p_t}+\Delta>0
\eq
and the strong energy conditions can be written as
\bq
C_6=2\,{\it p_t}+\Delta>0
\eq
where $\Delta=\sqrt {{{\rho}}^{2}+2\,{\it \rho}\,{\it p}+{{\it p}}^{2}-4\,{{
\it q}}^{2}}$.

If we assume that $q=0$ then we get the following inequalities:
\bq
C_1={\rho}+p >0, 
\eq
\bq
C_2={\rho}+ p_t>0,
\eq
\bq
C_3={\rho}>0,
\eq
\bq
C_4={\rho}-{\it p}>0
\eq
\bq
C_5={\rho}-p_t>0
\eq
\bq
C_6=2\,{\it p_t}+\rho+p>0.
\eq
Hereinafter, we will use the classification of types of matter given by Chan, 
da Silva and Villas 
da Rocha (2009) \cite{Chan2009}. If, for example, $C_1$ is not satisfied, but
$C_2$, $C_3$ and $C_6$ are satisfied, then we have a physical system with 
attractive phantom energy.

In the following we will study two particular cases, with suitable choices
for the parameter $\alpha$, in order to obtain analytical and manageable 
solutions for the proposed problem.

\section{Case $\alpha=-1$}

Let us now analyze a particular case where $\alpha=-1$ and $x_0=0$,
giving
\bq
\theta_l=\frac{2 r^2 (1+3 t r)^{\frac{2}{3}}+1+5 t r}{r(1+3 t r)^{\frac{5}{3}}},
\lb{thetal}
\eq
\bq
\theta_n=\frac{2 r^2 (1+3 t r)^{\frac{2}{3}}-1-5 t r}{r(1+3 t r)^{\frac{5}{3}}}.
\lb{thetan}
\eq

\begin{figure}
\vspace{.2in}
\centerline{\psfig{figure=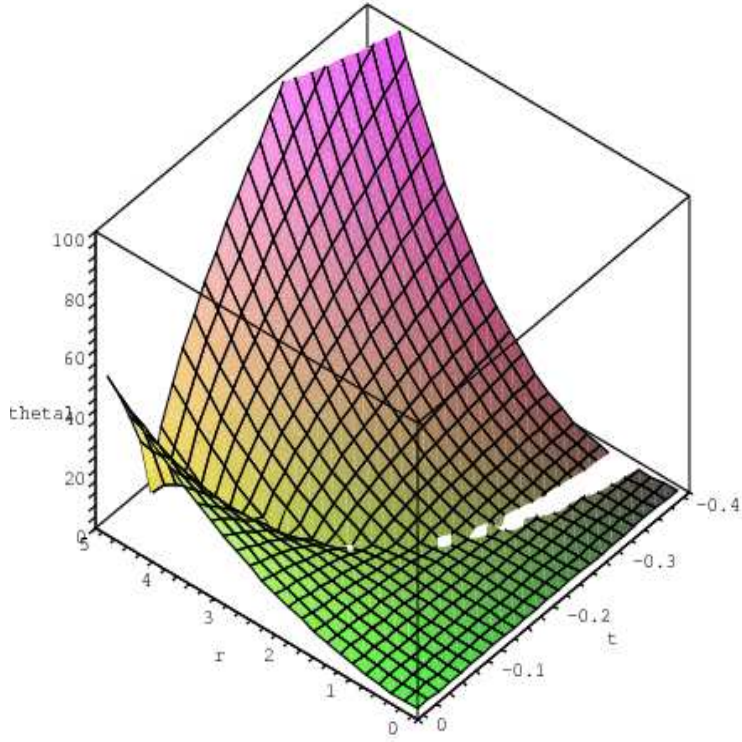,width=4.0 true in,height=4.0 true in}}
\caption{The outgoing null geodesic expansion $\theta_l$, for $\alpha=-1$ and $x_0=0$.}
\label{fig1}
\end{figure}
\begin{figure}
\vspace{.2in}
\centerline{\psfig{figure=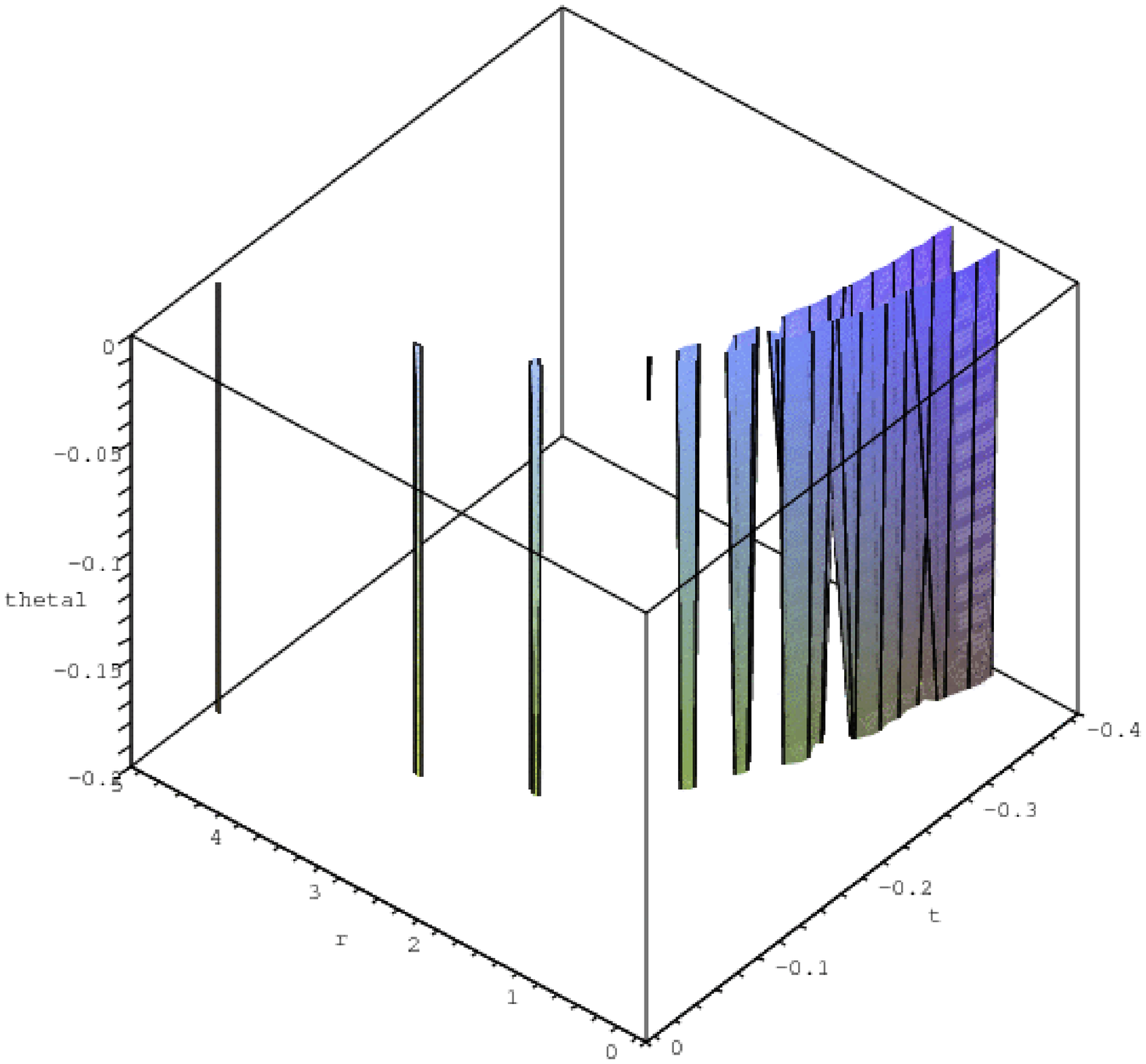,width=4.0 true in,height=4.0 true in}}
\caption{Zoom of the outgoing null geodesic expansion $\theta_l$, for $\alpha=-1$ and $x_0=0$.}
\label{fig2}
\end{figure}
\begin{figure}
\vspace{.2in}
\centerline{\psfig{figure=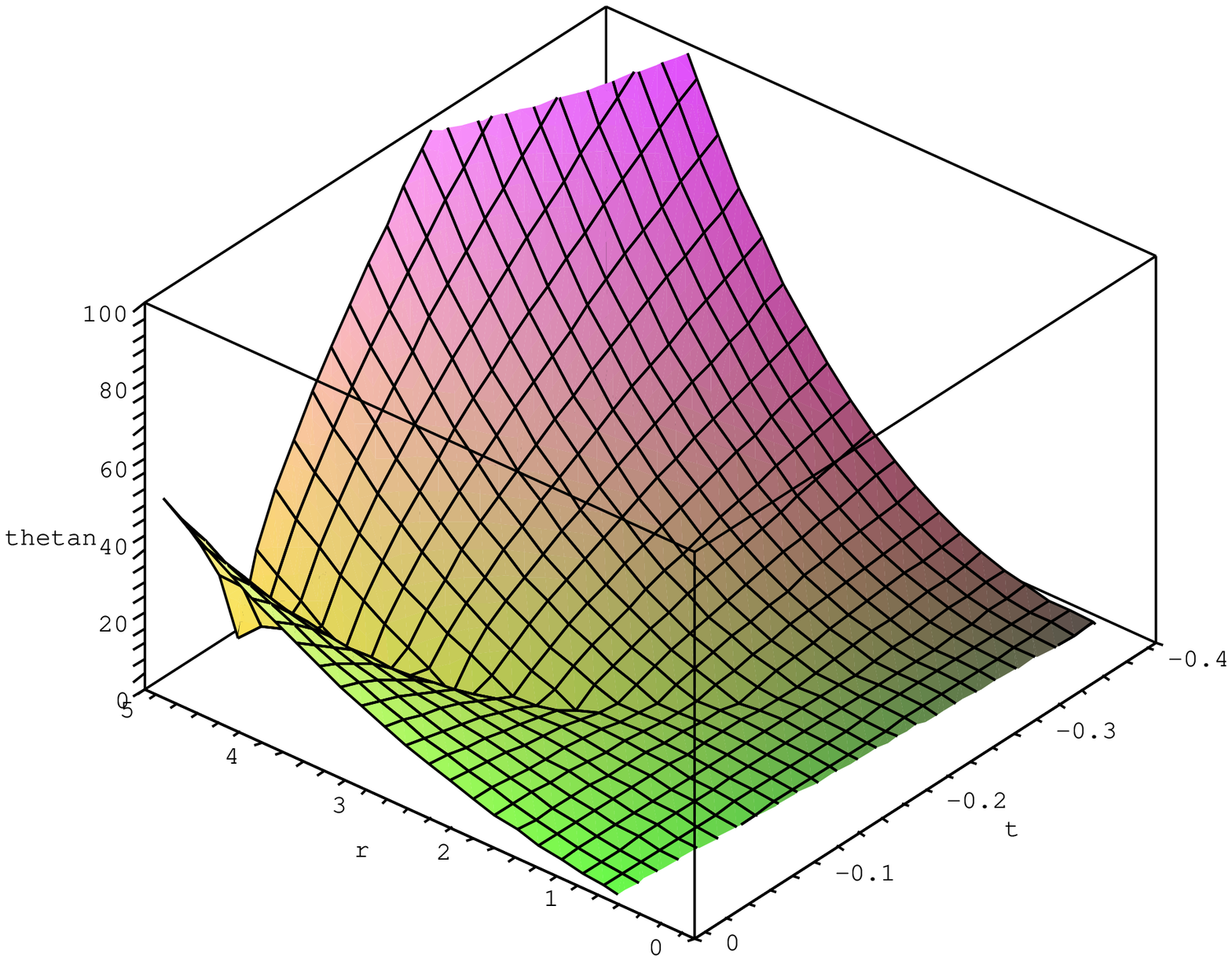,width=4.0 true in,height=4.0 true in}}
\caption{The ingoing null geodesic expansion $\theta_n$, for $\alpha=-1$ and $x_0=0$.}
\label{fig3}
\end{figure}
\begin{figure}
\vspace{.2in}
\centerline{\psfig{figure=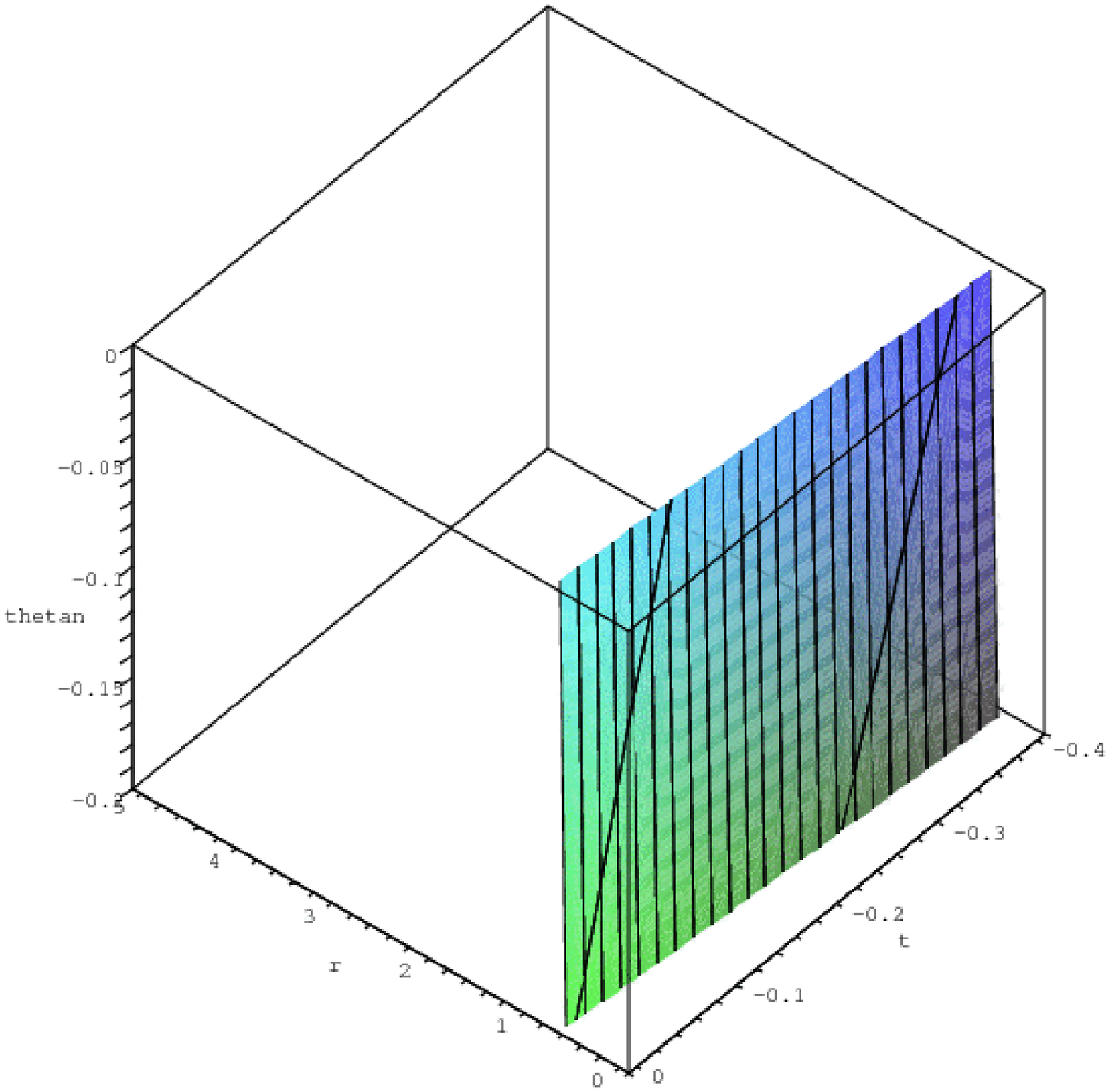,width=4.0 true in,height=4.0 true in}}
\caption{Zoom of the ingoing null geodesics expansion $\theta_n$, for $\alpha=-1$ and $x_0=0$.}
\label{fig4}
\end{figure}

Solving $\theta_l=0$ we get
\bqn
t_{ah} &=& \frac{1}{r}\left\{
\frac{2}{125}\left[3600r^{12}-1250r^6-1728r^{18}+250\sqrt{-16r^{18}+25r^{12}}\right]^{\frac{1}{3}} \right.-\nb \\
& &\left. \frac{\frac{125}{2}\left(\frac{32}{625}r^6-\frac{576}{15625}r^{12}\right)}{\left[3600r^{12}-1250r^6-1728r^{18}+250\sqrt{-16r^{18}+25r^{12}}\right]^{\frac{1}{3}}}-\frac{24}{125}r^6-\frac{1}{5} \right\}
\eqn
and solving $\theta_n=0$ we get
\bqn
t_n &=&\frac{1}{r} \left\{ \frac{2}{125}\left(3600r^{12}+1250r^6+1728r^{18}+250\sqrt{16r^{18}+25r^{12}}\right)^\frac{1}{3}- \right. \nb \\
& & \left. \frac{\frac{125}{2}(-\frac{32}{625}r^6-\frac{576}{15625}r^{12})}{\left[3600r^{12}+1250r^6+1728r^{18}+250\sqrt{16r^{18}+25r^{12}}\right]^{\frac{1}{3}}}+\frac{24}{125}r^6-\frac{1}{5} \right\}
\eqn

\begin{figure}
\vspace{.2in}
\centerline{\psfig{figure=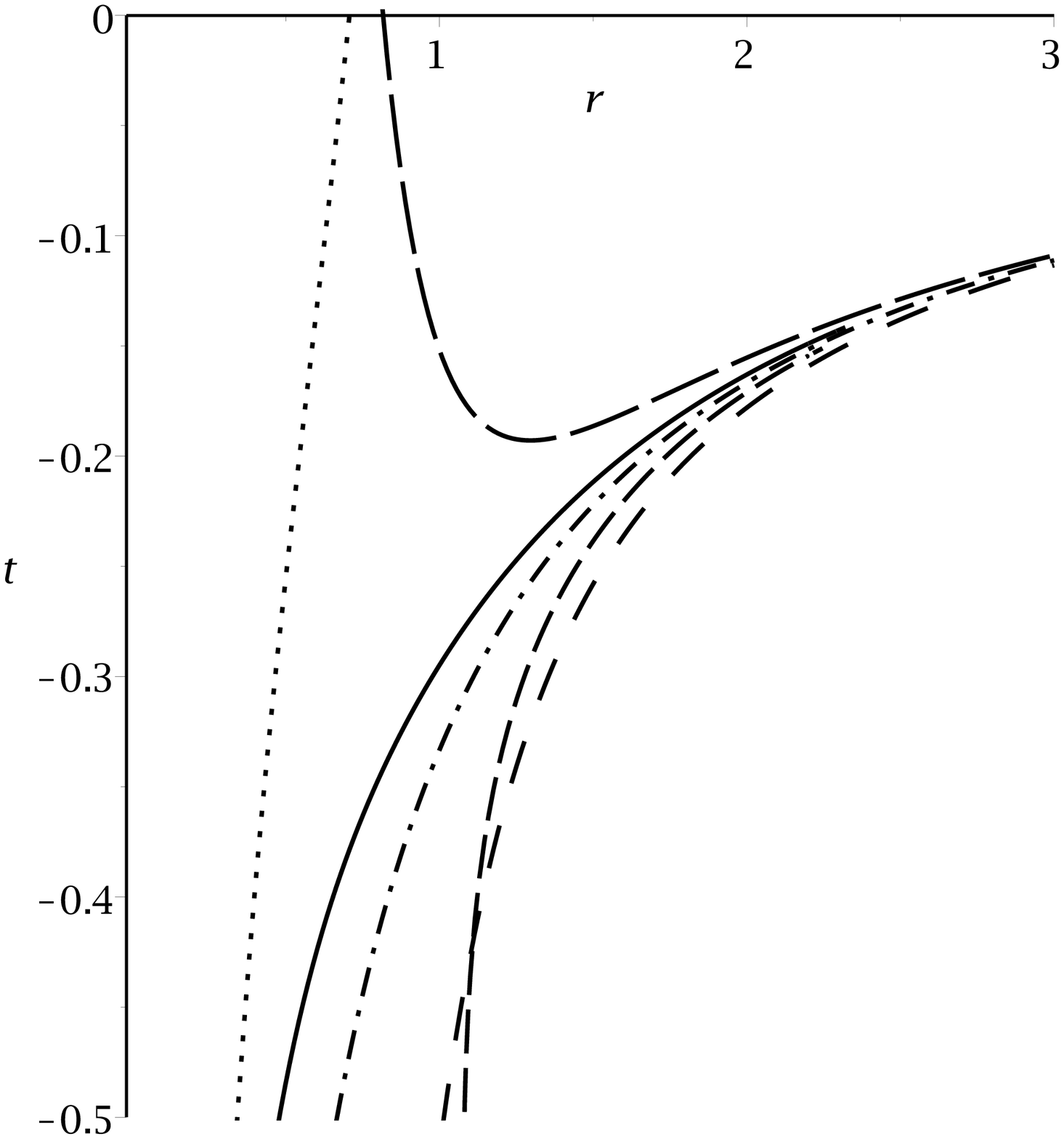,width=4.0 true in,height=4.0 true in}}
\caption{Apparent horizon $\theta_l=0$ (solid and dashed curves), for $\alpha=-1$ 
and $x_0=0$, $\theta_n=0$ curve (dotted curve), singularity curve (dot-dashed curve), $C_1$ energy condition curves 
(long-dashed and space-dashed curves) where the energy condition $C_1$ is not fulfilled.}
\label{fig5}
\end{figure}

\begin{figure}
\vspace{.2in}
\centerline{\psfig{figure=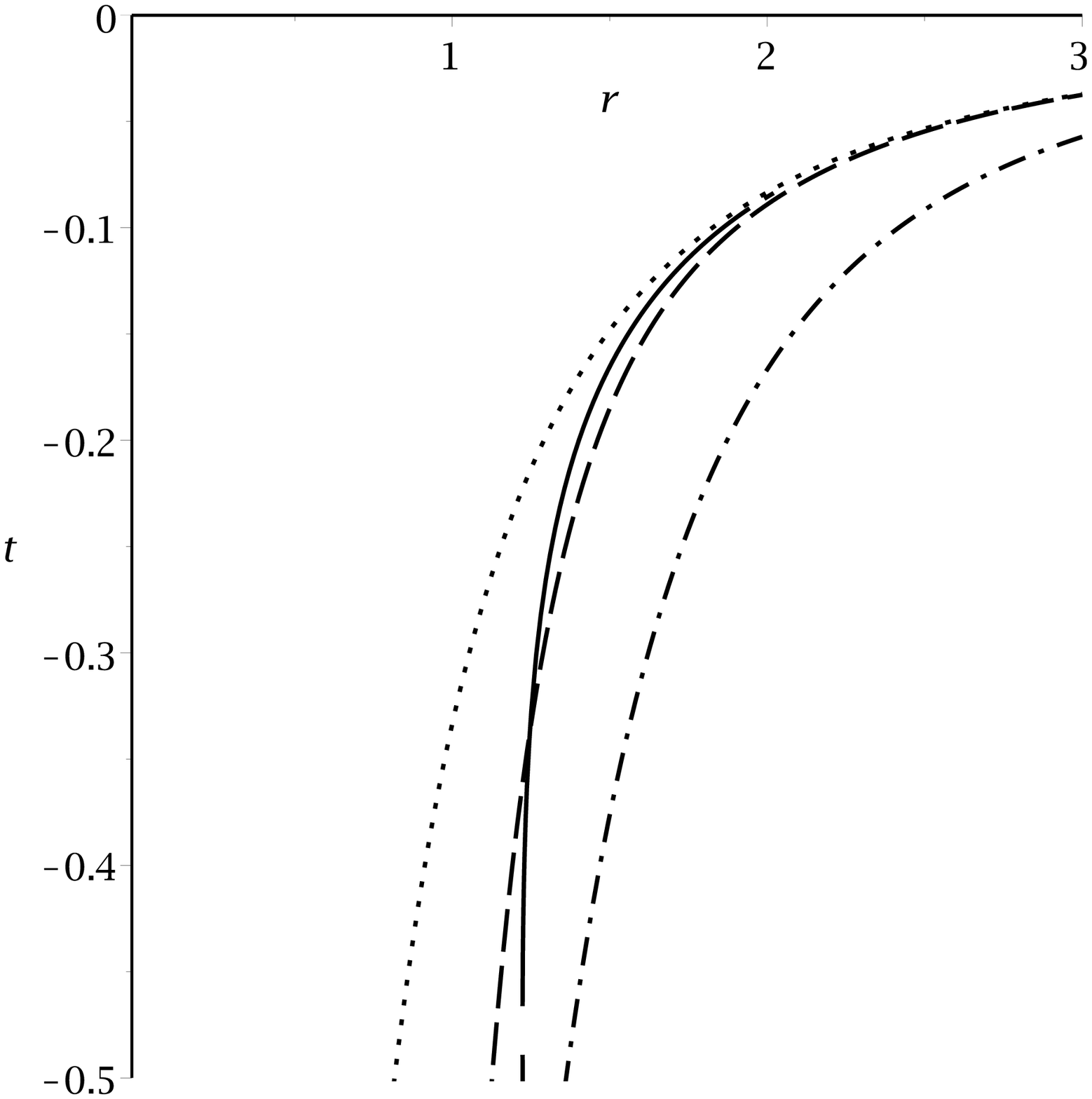,width=4.0 true in,height=4.0 true in}}
\caption{Apparent horizon (solid curve)  $\theta_l=0$, for $\alpha=-1$ and 
$x_0=0$, singularity curve (dotted curve) and $C_1$ energy condition curve 
(dashed curve). The time $t=t_c \approx -0.14$ (where $r_c \approx 2.35$) 
denotes the moment 
when the apparent horizon curve disappears. The dotted curve represents also 
$R=0$ and the dot-dashed curve denotes $R=1.2$.}
\label{fig5a}
\end{figure}

In order to have a black hole we should have $\theta_n$ horizon being
interior to the $\theta_l$ horizon, which is the case (see Figure \ref{fig5a}).
However, another condition to have black hole is to have $\theta_n<0$, while $\theta_l>0$, in the exterior
to the $\theta_n$ horizon.  Since this horizon is located inside the
singularity, we do not have to analyze it.
The Figure \ref{fig5a} shows that the apparent horizon does not cover the
singularity for a radius greater than $r_c$,  then the structure characterizes a 
naked singularity
formation. For the radius $ r < r_c$, there is a region of exotic matter,
localized before the apparent horizon.
The apparent  horizon disappears for the radius $r > r_c$,
but the singularity remains, then it seems to exist a naked singularity there.
The disappearance of the apparent horizon may be related to the presence of 
exotic matter.

The density, radial and tangential pressure, and the heat flow are given by
\bq
\rho=
4\,{\frac {9\,{r}^{4}\sqrt [3]{{\it \delta_1}}t+3\,{r}^{3}\sqrt [3]{{
\it \delta_1}}-4\,{t}^{2}r-t}{{{\it \delta_1}}^{10/3}r}},
\eq
\bq
p_r=
4\,{\frac {t \left( 1+4\,tr \right) }{{{\it \delta_1}}^{10/3}r}},
\eq
\bq
p_t=
2\,{\frac {t}{{{\it \delta_1}}^{10/3}r}},
\eq
\bq
q=
-4\,{{\it \delta_1}}^{-8/3},
\eq
and
where $\delta_1=1+3 t r$.

The energy conditions are given by
\bq
C_1=
4\, \left( 3\, \left| {\frac {{r}^{2} \left( 1+3\,tr \right) }{{{\it 
\delta_1}}^{3}}} \right|  \left(  \left| {\it \delta_1} \right|  \right) ^
{8/3}-2 \right)  \left(  \left| {\it \delta_1} \right|  \right) ^{-8/3}>0,
\eq
\bq
C_2=
4\,{\frac {9\,{r}^{4}\sqrt [3]{{\it \delta_1}}t+3\,{r}^{3}\sqrt [3]{{
\it \delta_1}}-4\,{t}^{2}r-t}{{{\it \delta_1}}^{10/3}r}}-4\,{\frac {t
 \left( 1+4\,tr \right) }{{{\it \delta_1}}^{10/3}r}}+4\,{\frac {t}{{{
\it \delta_1}}^{10/3}r}}+\Delta_1>0,
\eq
\bq
C_3=
4\,{\frac {9\,{r}^{4}\sqrt [3]{{\it \delta_1}}t+3\,{r}^{3}\sqrt [3]{{
\it \delta_1}}-4\,{t}^{2}r-t}{{{\it \delta_1}}^{10/3}r}}-4\,{\frac {t
 \left( 1+4\,tr \right) }{{{\it \delta_1}}^{10/3}r}}+\Delta_1>0
\eq
\bq
C_4=
4\,{\frac {9\,{r}^{4}\sqrt [3]{{\it \delta_1}}t+3\,{r}^{3}\sqrt [3]{{
\it \delta_1}}-4\,{t}^{2}r-t}{{{\it \delta_1}}^{10/3}r}}-4\,{\frac {t
 \left( 1+4\,tr \right) }{{{\it \delta_1}}^{10/3}r}}>0
\eq
\bq
C_5=
4\,{\frac {9\,{r}^{4}\sqrt [3]{{\it \delta_1}}t+3\,{r}^{3}\sqrt [3]{{
\it \delta_1}}-4\,{t}^{2}r-t}{{{\it \delta_1}}^{10/3}r}}-4\,{\frac {t
 \left( 1+4\,tr \right) }{{{\it \delta_1}}^{10/3}r}}-4\,{\frac {t}{{{
\it \delta_1}}^{10/3}r}}+\Delta_1>0
\eq
\bq
C_6=
4\,{\frac {t}{{{\it \delta_1}}^{10/3}r}}+\Delta_1>0
\eq
where 
\bq
\Delta_1=\Delta=
4\,\sqrt {-{\frac {-81\,{r}^{6}{t}^{2}-54\,{r}^{5}t-9\,{r}^{4}+4\,{{
\it \delta_1}}^{2/3}}{{{\it \delta_1}}^{6}}}}
\eq

\begin{figure}
\vspace{.2in}
\centerline{  \psfig{figure=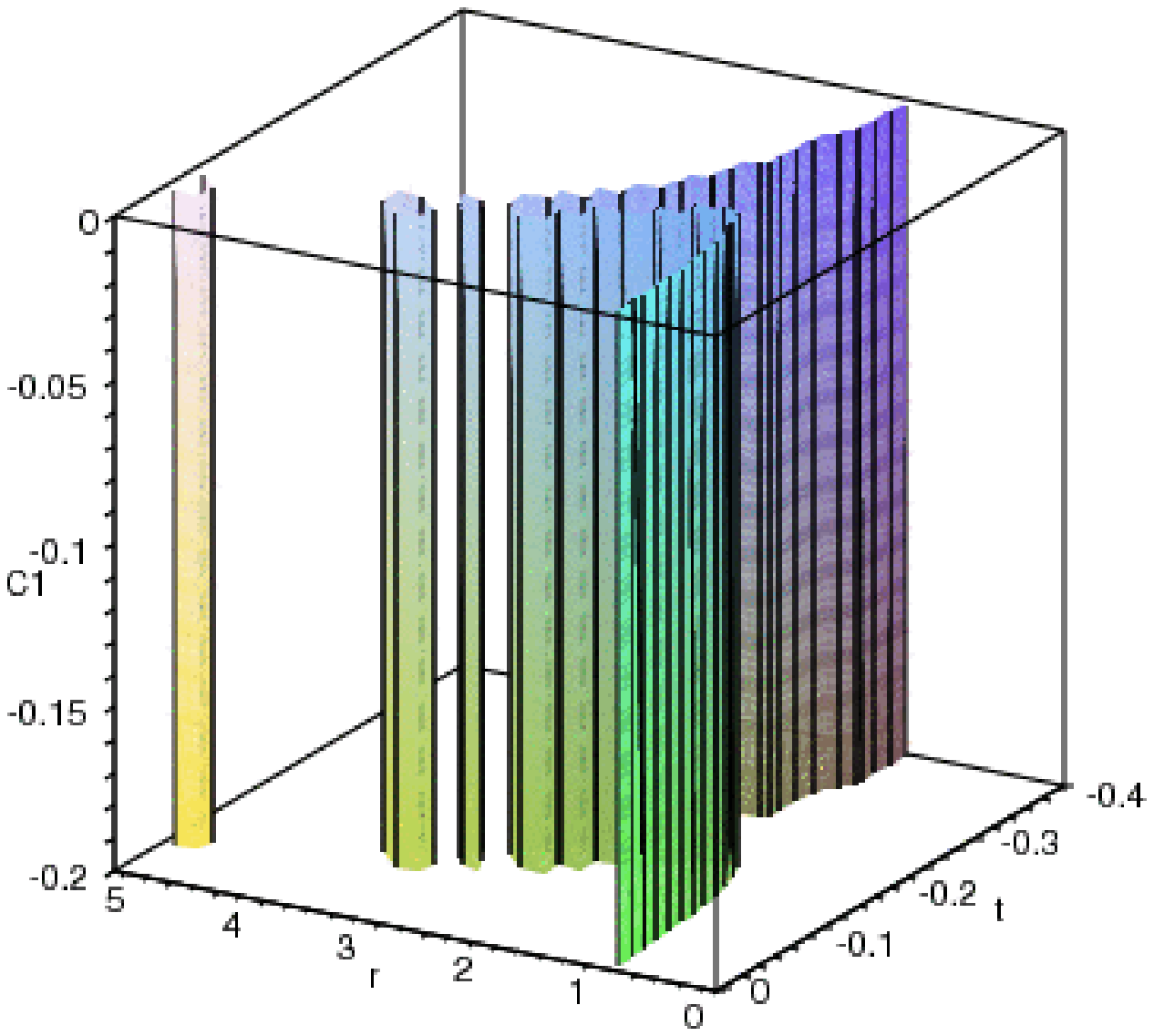,width=3.0truein,height=3.0truein}
\hskip .1in   \psfig{figure=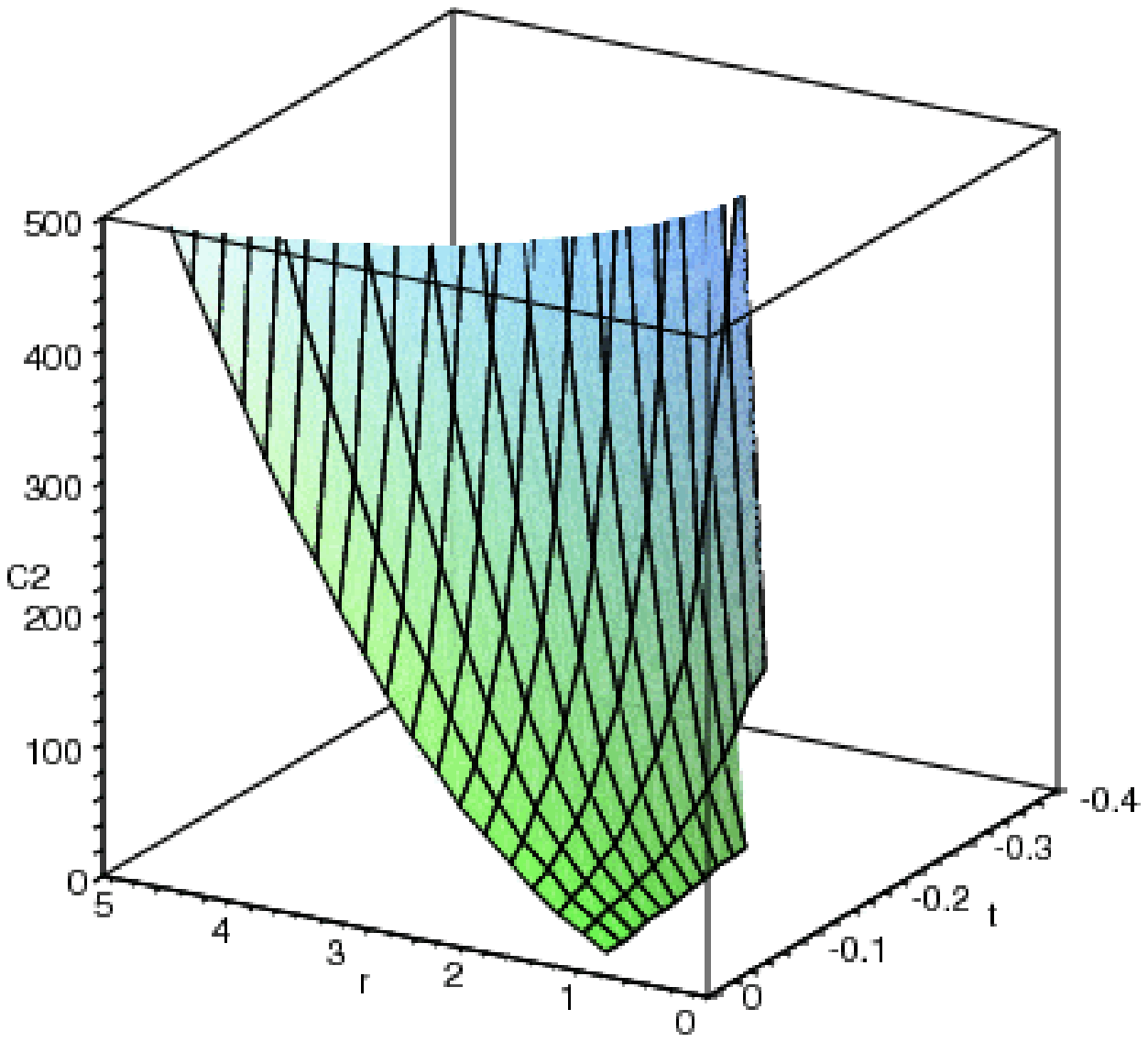,width=3.0truein,height=3.0truein}
\hskip .5in}
\centerline{  \psfig{figure=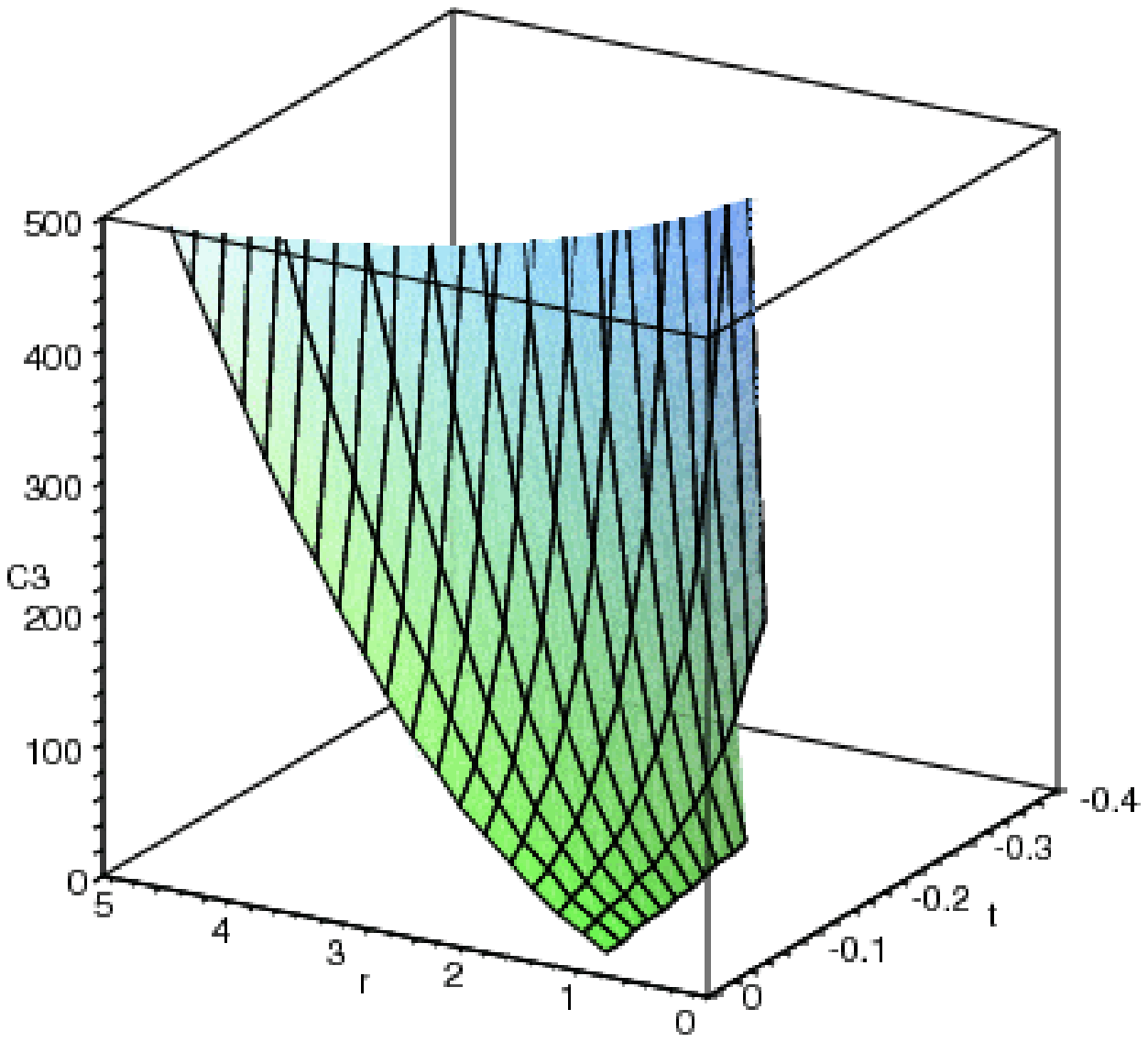,width=3.0truein,height=3.0truein}
\hskip .1in   \psfig{figure=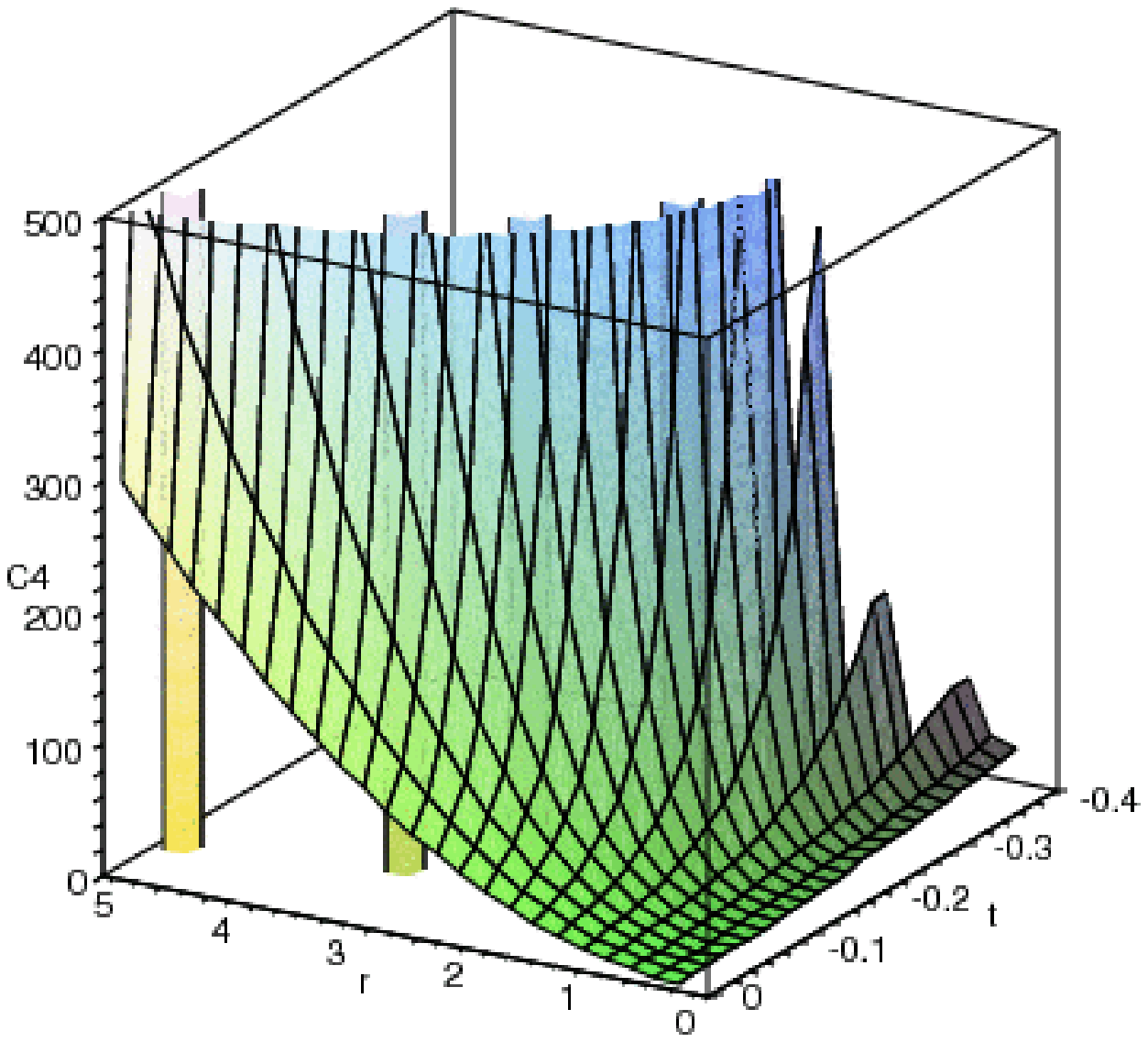,width=3.0truein,height=3.0truein}
\hskip .5in}
\centerline{  \psfig{figure=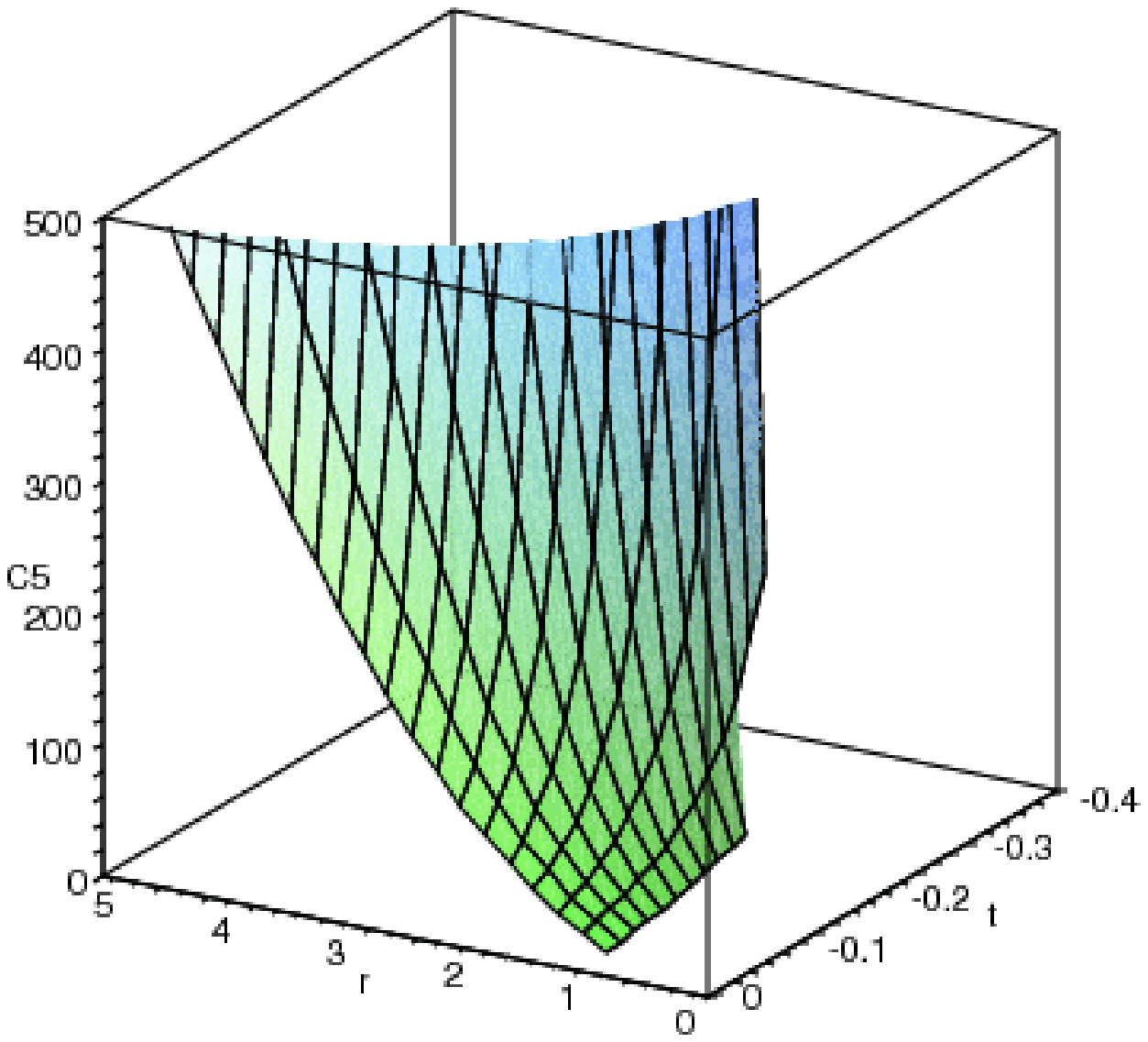,width=3.0truein,height=3.0truein}
\hskip .1in   \psfig{figure=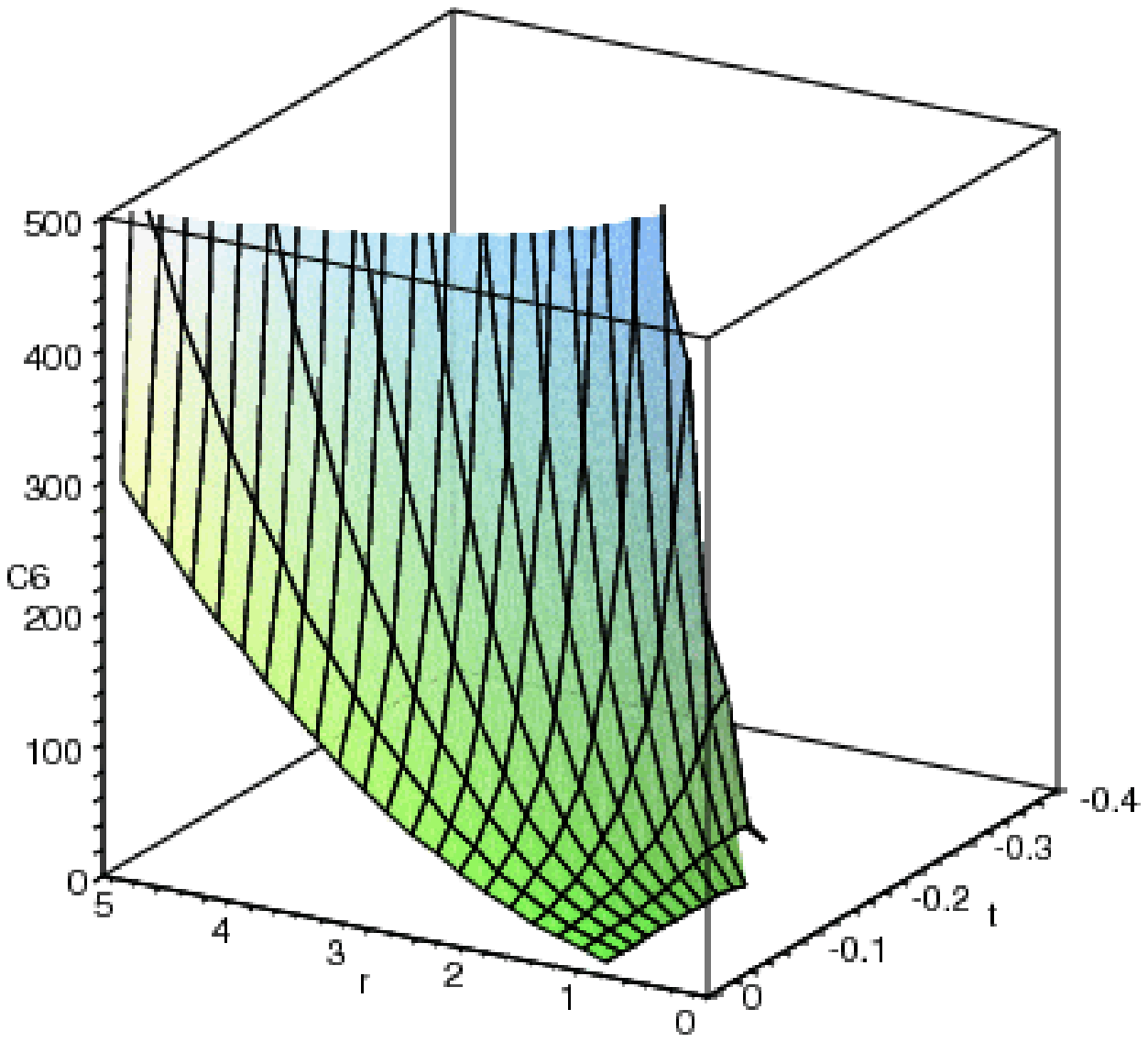,width=3.0truein,height=3.0truein}
\hskip .5in}  \caption{The energy conditions for $\alpha=-1$ and $x_0=0$.}
\label{condalpham1}
\end{figure}

\begin{figure}
\vspace{.2in}
\centerline{  \psfig{figure=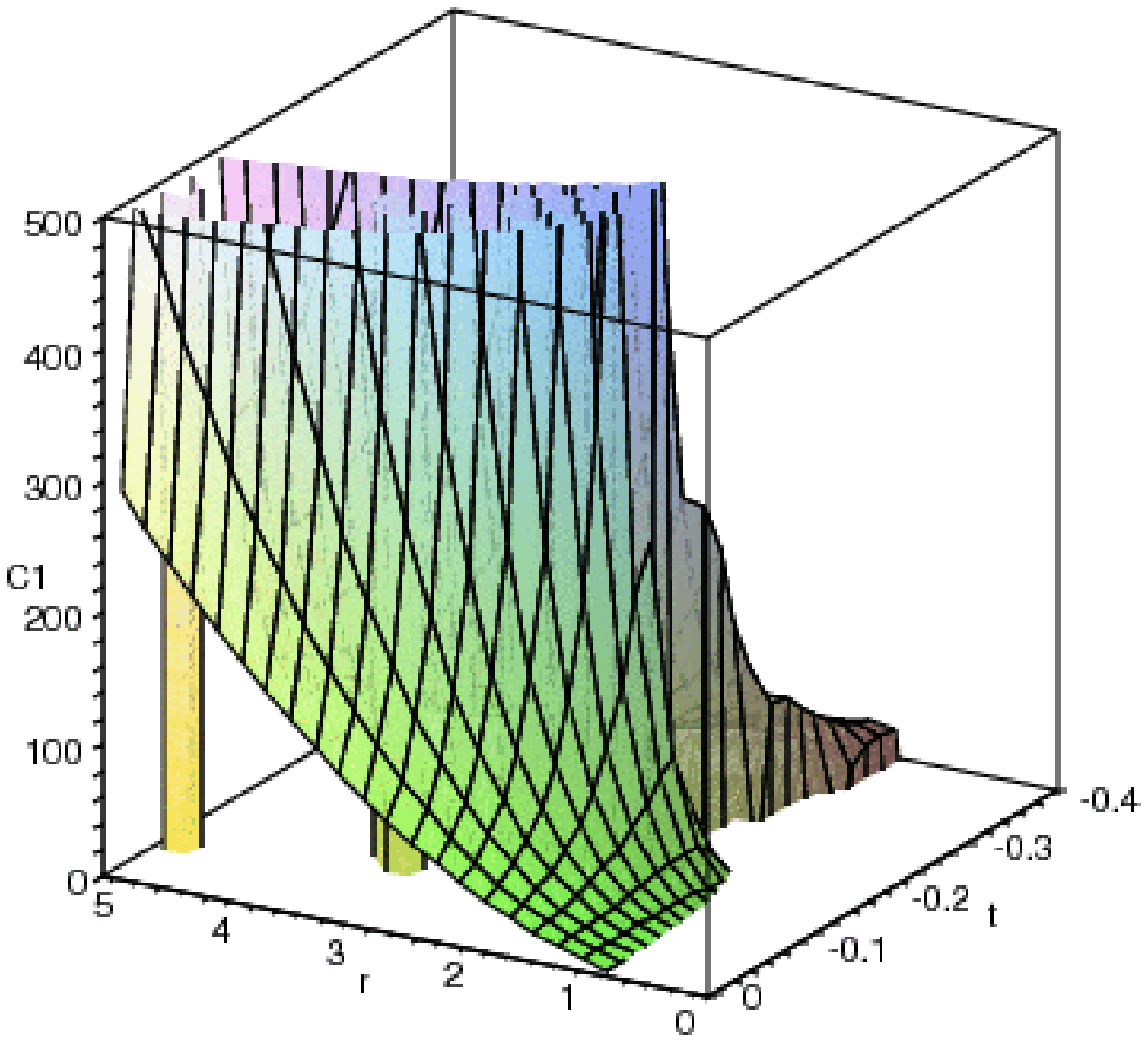,width=3.0truein,height=3.0truein}
\hskip .1in   \psfig{figure=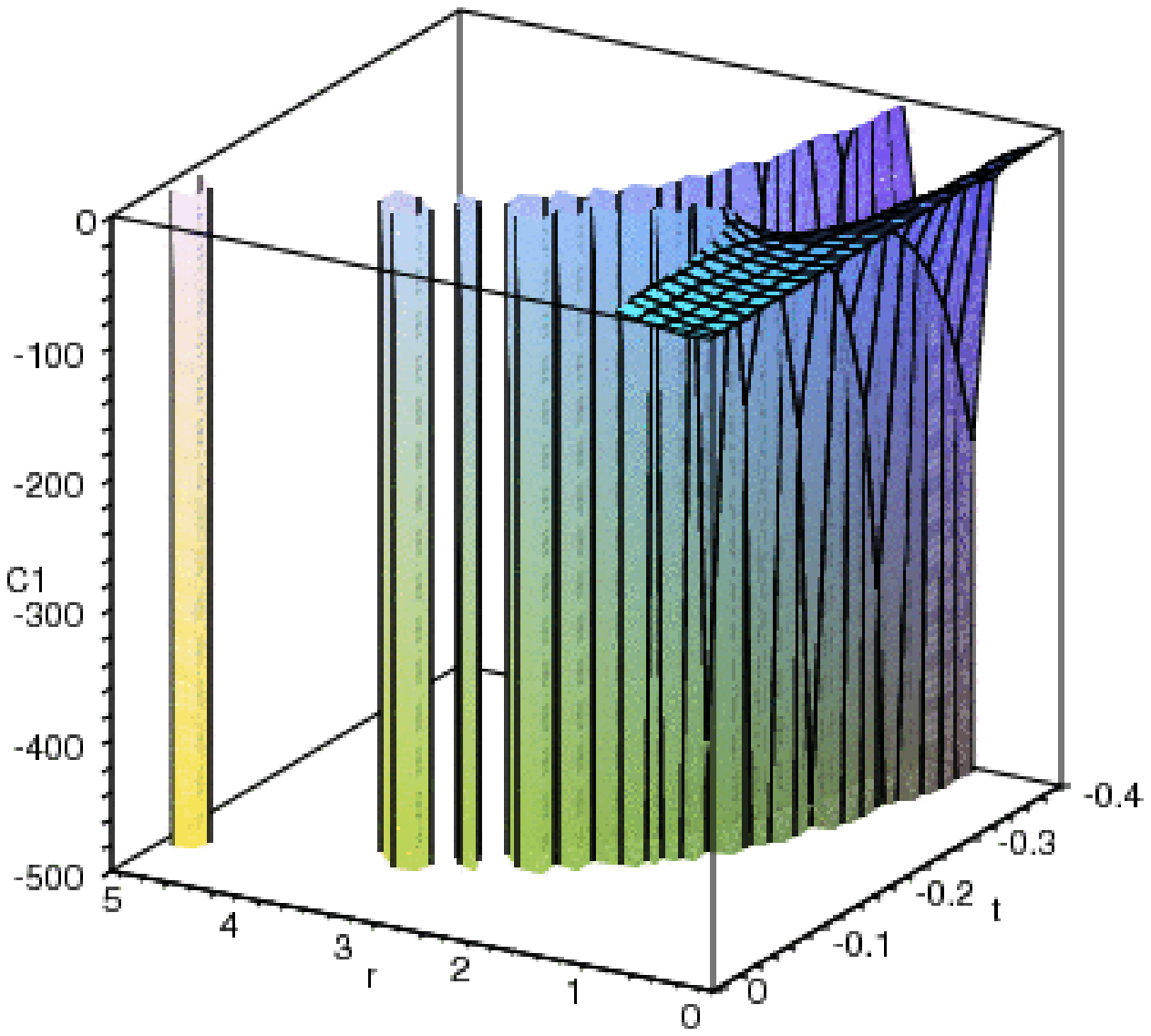,width=3.0truein,height=3.0truein}
\hskip .5in}  \caption{The positive and negative part of the energy condition 
$C1$ for $\alpha=-1$ and $x_0=0$.}
\label{condalpham1a}
\end{figure}

We can see in Figure \ref{condalpham1} that all the energy conditions are 
satisfied except the Condition 1 ($C_1$)
in the neighborhood of 
the curves where $C_1=0$ 
(see also Figure \ref{condalpham1a}).

\section{Case $\alpha=-2$}

Let us now analyze a particular case where $\alpha=-2$ and $x_0=0$, giving
\bq
\theta_l=\frac{2 r^3 (1+3 t r^2)^{\frac{2}{3}}+1+7 t r^2}{r(1+3 t r^2)^{\frac{5}{3}}},
\lb{thetal_a}
\eq
\bq
\theta_n=\frac{2 r^2 (1+3 t r^2)^{\frac{2}{3}}-1-7 t r^2}{r(1+3 t r^2)^{\frac{5}{3}}}.
\lb{thetan_a}
\eq

\begin{figure}
\vspace{.2in}
\centerline{\psfig{figure=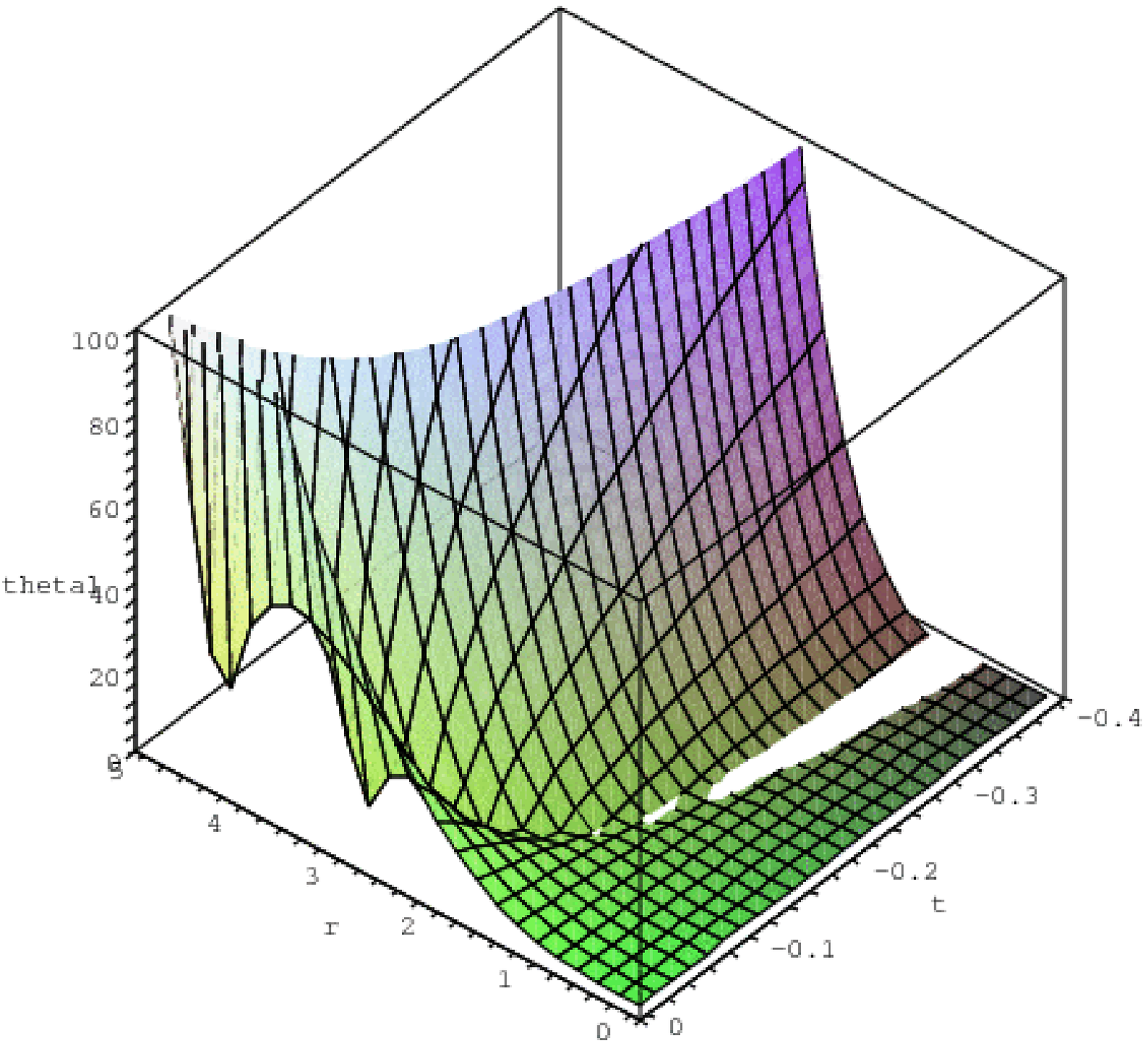,width=4.0 true in,height=4.0 true in}}
\caption{The outgoing null geodesic expansion $\theta_l$, for $\alpha=-2$ and $x_0=0$.}
\label{fig1_a}
\end{figure}
\begin{figure}
\vspace{.2in}
\centerline{\psfig{figure=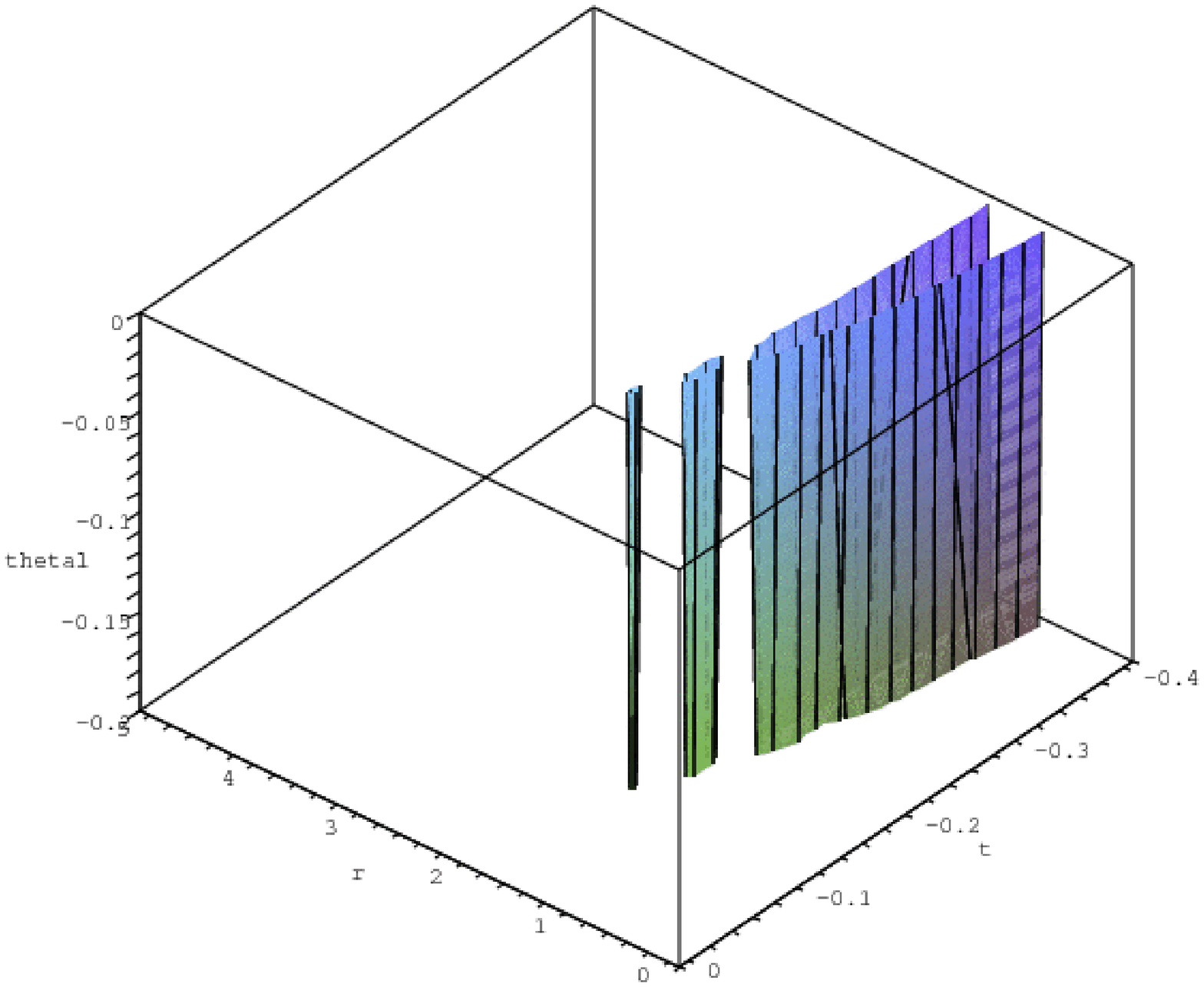,width=4.0 true in,height=4.0 true in}}
\caption{Zoom of the outgoing null geodesic expansion $\theta_l$, for $\alpha=-2$ and $x_0=0$.}
\label{fig2_a}
\end{figure}
\begin{figure}
\vspace{.2in}
\centerline{\psfig{figure=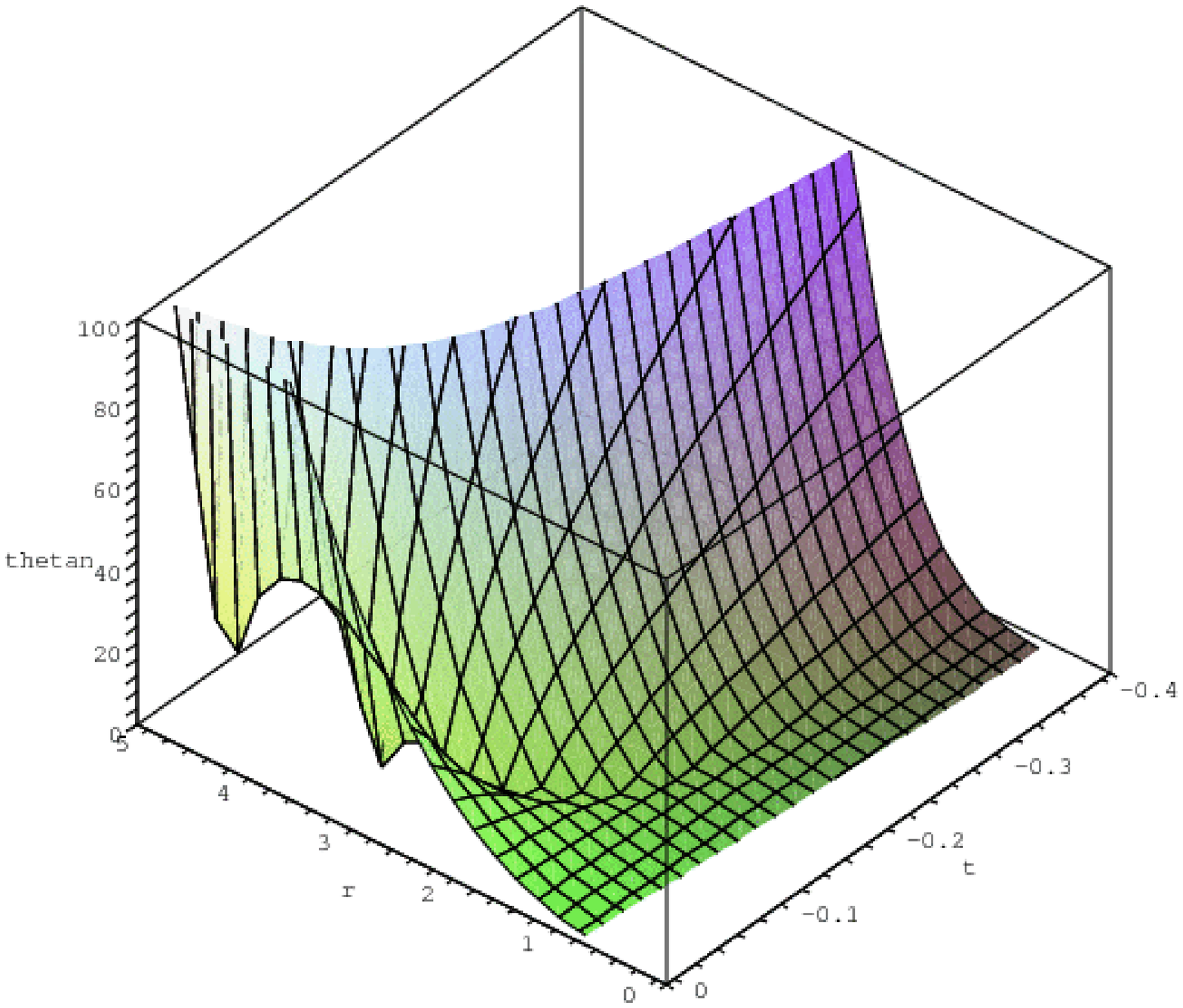,width=4.0 true in,height=4.0 true in}}
\caption{The ingoing null geodesic expansion $\theta_n$, for $\alpha=-2$ and $x_0=0$.}
\label{fig3_a}
\end{figure}
\begin{figure}
\vspace{.2in}
\centerline{\psfig{figure=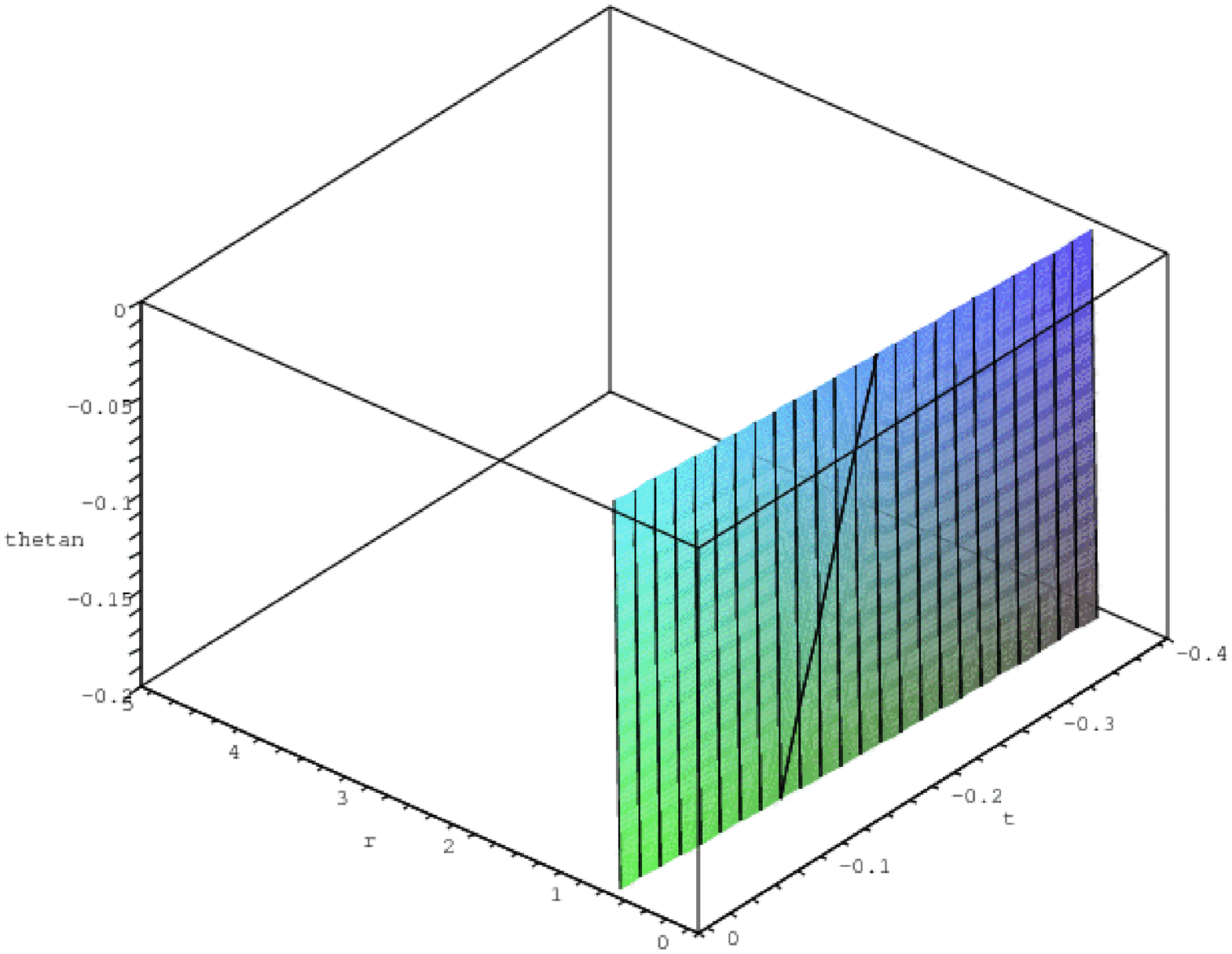,width=4.0 true in,height=4.0 true in}}
\caption{Zoom of the ingoing null geodesics expansion $\theta_n$, for $\alpha=-2$ and $x_0=0$.}
\label{fig4_a}
\end{figure}

Solving $\theta_l=0$ we get
\bqn
t_{ah} &=& \frac{1}{r^2}\left\{
\frac{4}{343}\left[1764r^{18}-2401r^9-216r^{27}+343\sqrt{-8r^{27}+49r^{18}}\right]^{\frac{1}{3}} \right.-\nb \\
& &\left. \frac{\frac{343}{4}\left(\frac{64}{2401}r^9-\frac{576}{117649}r^{18}\right)}{\left[1764r^{18}-2401r^9-216r^{27}+343\sqrt{-8r^{27}+49r^{18}}\right]^{\frac{1}{3}}}-\frac{24}{343}r^9-\frac{1}{7} \right\}
\eqn
and solving $\theta_n=0$ we get
\bqn
t_{n} &=& \frac{1}{r^2}\left\{
\frac{4}{343}\left[1764r^{18}+2401r^9-216r^{27}+343\sqrt{8r^{27}+49r^{18}}\right]^{\frac{1}{3}} \right.-\nb \\
& &\left. \frac{\frac{343}{4}\left(\frac{64}{2401}r^9-\frac{576}{117649}r^{18}\right)}{\left[1764r^{18}+2401r^9+216r^{27}+343\sqrt{8r^{27}+49r^{18}}\right]^{\frac{1}{3}}}-\frac{24}{343}r^9-\frac{1}{7} \right\}
\eqn

\begin{figure}
\vspace{.2in}
\centerline{\psfig{figure=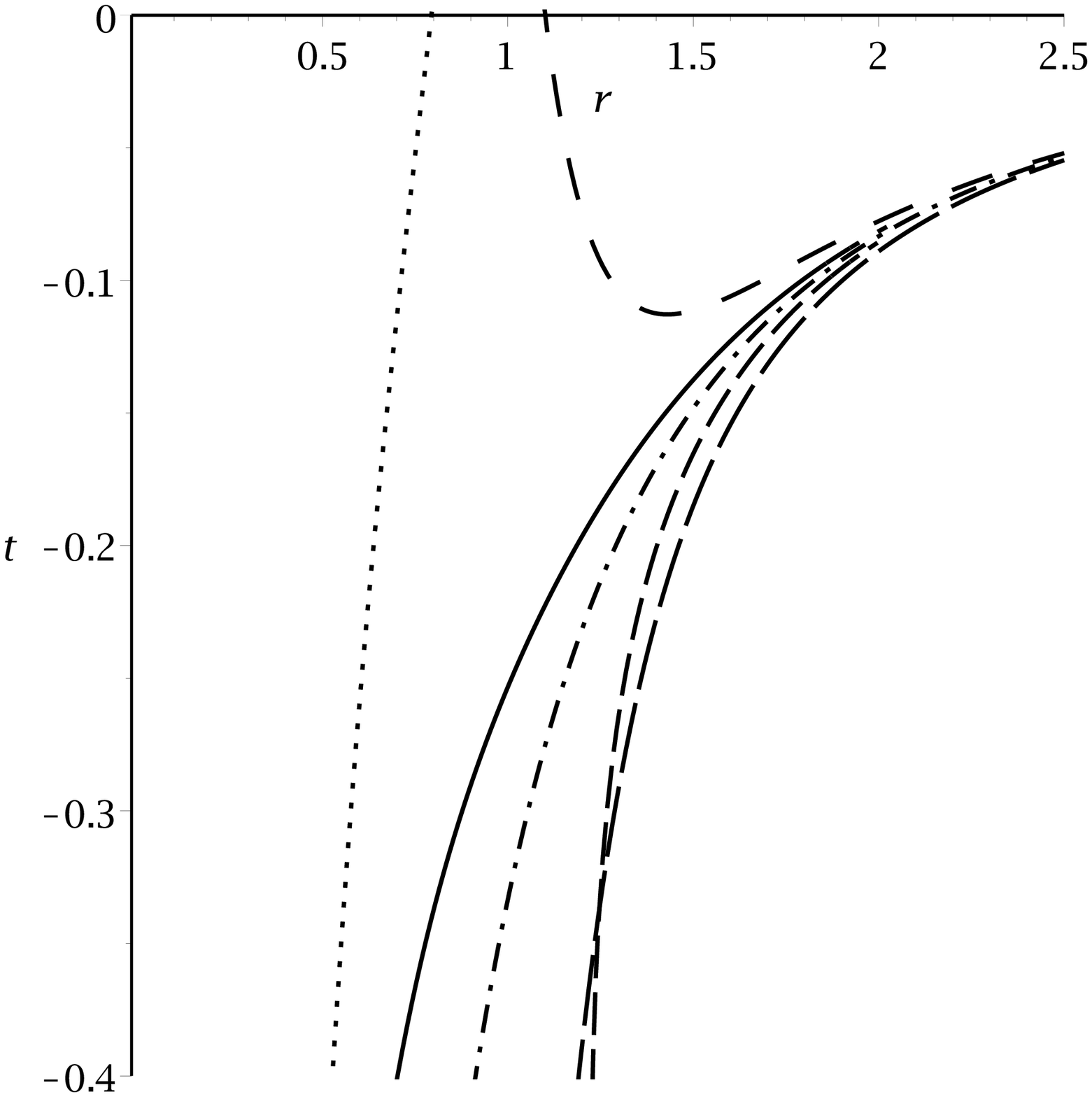,width=4.0 true in,height=4.0 true in}}
\caption{Apparent horizon $\theta_l=0$ (solid and dashed curves), for $\alpha=-2$ 
and $x_0=0$, $\theta_n=0$ curve (dotted curve), singularity curve (dot-dashed curve), $C_1$ energy condition curves 
(long-dashed and space-dashed curves) where the energy condition $C_1$ is not fulfilled.}
\label{fig5_a}
\end{figure}

\begin{figure}
\vspace{.2in}
\centerline{\psfig{figure=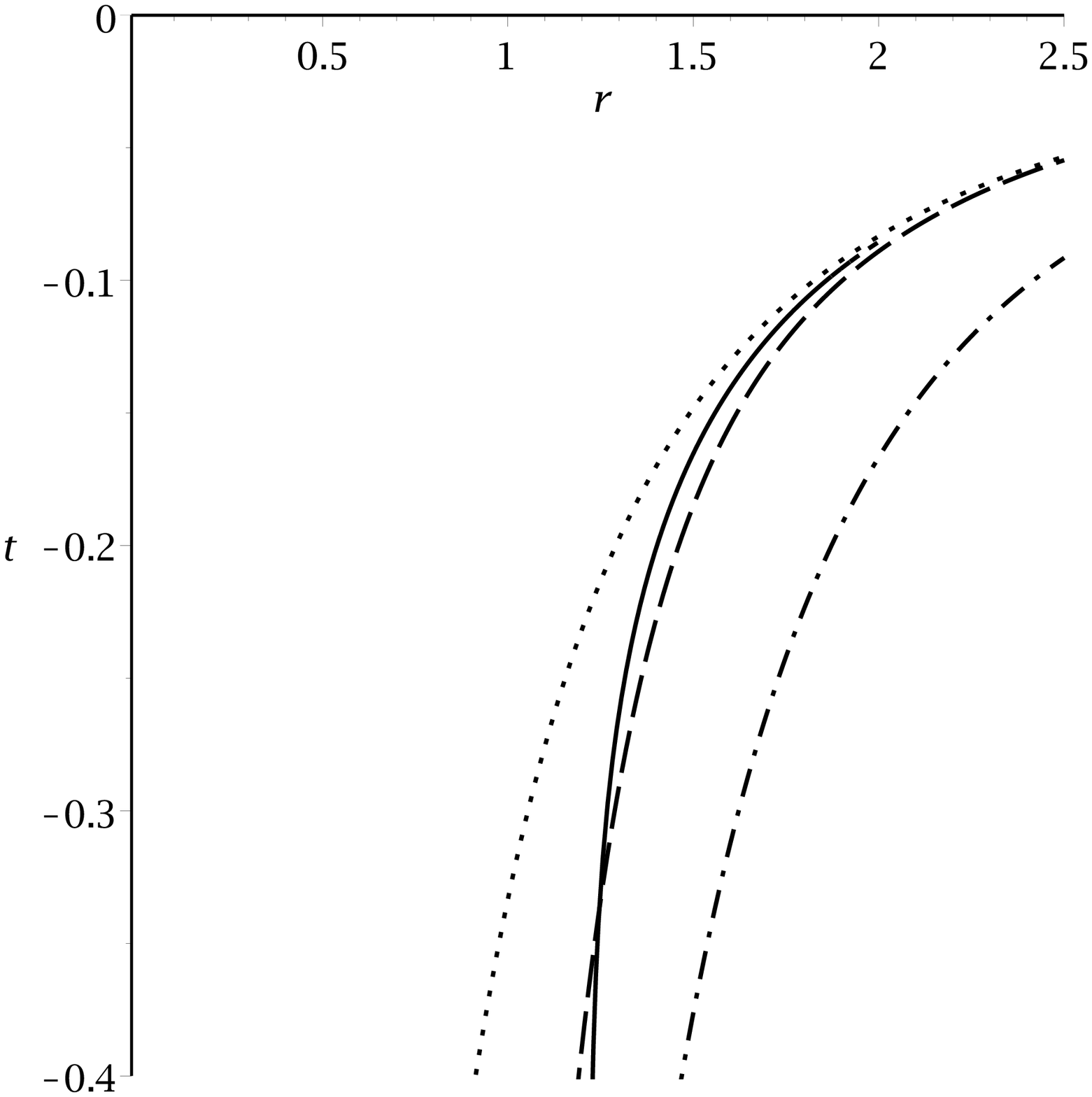,width=4.0 true in,height=4.0 true in}}
\caption{Apparent horizon (solid curve)  $\theta_l=0$, for $\alpha=-2$ and 
$x_0=0$, singularity curve (dotted curve) and $C_1$ energy condition curve 
(dashed curve). The time  $t=t_c \approx -0.17$ (where $r_c \approx 2.0$) 
denotes the time when the apparent horizon curve disappears.
The dotted curve represents also $R=0$ and the dot-dashed curve
denotes $R=2$.}
\label{fig5_ab}
\end{figure}

As in the case $\alpha=-1$, in order to have a black hole we should have 
that $\theta_n$ horizon being
interior to the $\theta_l$ horizon, which is the case (see Figure 
\ref{fig5_ab}).
However, another condition to have a black hole is to have $\theta_n<0$ exterior
to the $\theta_n$ horizon. Again, since this horizon is located inside the
singularity, we do not have to analyze it. 
Again, the Figure \ref{fig5_ab} shows that the apparent horizon covers the
singularity for a radius greater than $r_c$, then the structure characterizes a 
naked singularity
formation. For the radius $ r < r_c$, there is a region of exotic matter,
localized before the apparent horizon.
The apparent  horizon disappears for the radius $r > r_c$,
but the singularity remains, then it seems that there is a naked singularity there.

The density, radial and tangential pressure, and the heat flow are given by
\bq
\rho=
4\,{\frac {9\,{r}^{6}\sqrt [3]{{\it \delta_2}}t+3\,{r}^{4}\sqrt [3]{{
\it \delta_2}}-10\,{t}^{2}{r}^{2}-2\,t}{{{\it \delta_2}}^{10/3}}},
\eq
\bq
p_r=
8\,{\frac {t \left( 1+5\,{r}^{2}t \right) }{{{\it \delta_2}}^{10/3}}},
\eq
\bq
p_t=
8\,{\frac {t}{{{\it \delta_2}}^{10/3}}},
\eq
\bq
q=
-8\,{\frac {r}{{{\it \delta_2}}^{8/3}}},
\eq
where $\delta_2=1+3 t r^2$.

The energy conditions are given by
\bq
C_1=
12\, \left| {\frac {{r}^{4}}{{{\it 
\delta_2}}^{2}}} \right| -16\, \left| {\frac {r}{{{\it \delta_2}}^{8/3}}}
 \right|>0
\eq
\bq
C_2=
4\,{\frac {9\,{r}^{6}\sqrt [3]{{\it \delta_2}}t+3\,{r}^{4}\sqrt [3]{{
\it \delta_2}}-10\,{t}^{2}{r}^{2}-2\,t}{{{\it \delta_2}}^{10/3}}}-8\,{
\frac {t \left( 1+5\,{r}^{2}t \right) }{{{\it \delta_2}}^{10/3}}}+16\,{
\frac {t}{{{\it \delta_2}}^{10/3}}}+\Delta_2>0
\eq
\bq
C_3=
4\,{\frac {9\,{r}^{6}\sqrt [3]{{\it \delta_2}}t+3\,{r}^{4}\sqrt [3]{{
\it \delta_2}}-10\,{t}^{2}{r}^{2}-2\,t}{{{\it \delta_2}}^{10/3}}}-8\,{
\frac {t \left( 1+5\,{r}^{2}t \right) }{{{\it \delta_2}}^{10/3}}}+\Delta_2>0
\eq
\bq
C_4=
4\,{\frac {9\,{r}^{6}\sqrt [3]{{\it \delta_2}}t+3\,{r}^{4}\sqrt [3]{{
\it \delta_2}}-10\,{t}^{2}{r}^{2}-2\,t}{{{\it \delta_2}}^{10/3}}}-8\,{
\frac {t \left( 1+5\,{r}^{2}t \right) }{{{\it \delta_2}}^{10/3}}}>0
\eq
\bq
C_5=
4\,{\frac {9\,{r}^{6}\sqrt [3]{{\it \delta_2}}t+3\,{r}^{4}\sqrt [3]{{
\it \delta_2}}-10\,{t}^{2}{r}^{2}-2\,t}{{{\it \delta_2}}^{10/3}}}-8\,{
\frac {t \left( 1+5\,{r}^{2}t \right) }{{{\it \delta_2}}^{10/3}}}-16\,{
\frac {t}{{{\it \delta_2}}^{10/3}}}+\Delta_2>0
\eq
\bq
C_6=
16\,{\frac {t}{{{\it \delta_2}}^{10/3}}}+\Delta_2>0
\eq
where
\bq
\Delta_2=\Delta=
4\,\sqrt {-{\frac {{r}^{2} \left( -81\,{r}^{10}{t}^{2}-54\,{r}^{8}t-9
\,{r}^{6}+8\,{{\it \delta_2}}^{2/3} \right) }{{{\it \delta_2}}^{6}}}}.
\eq

\begin{figure}
\vspace{.2in}
\centerline{  \psfig{figure=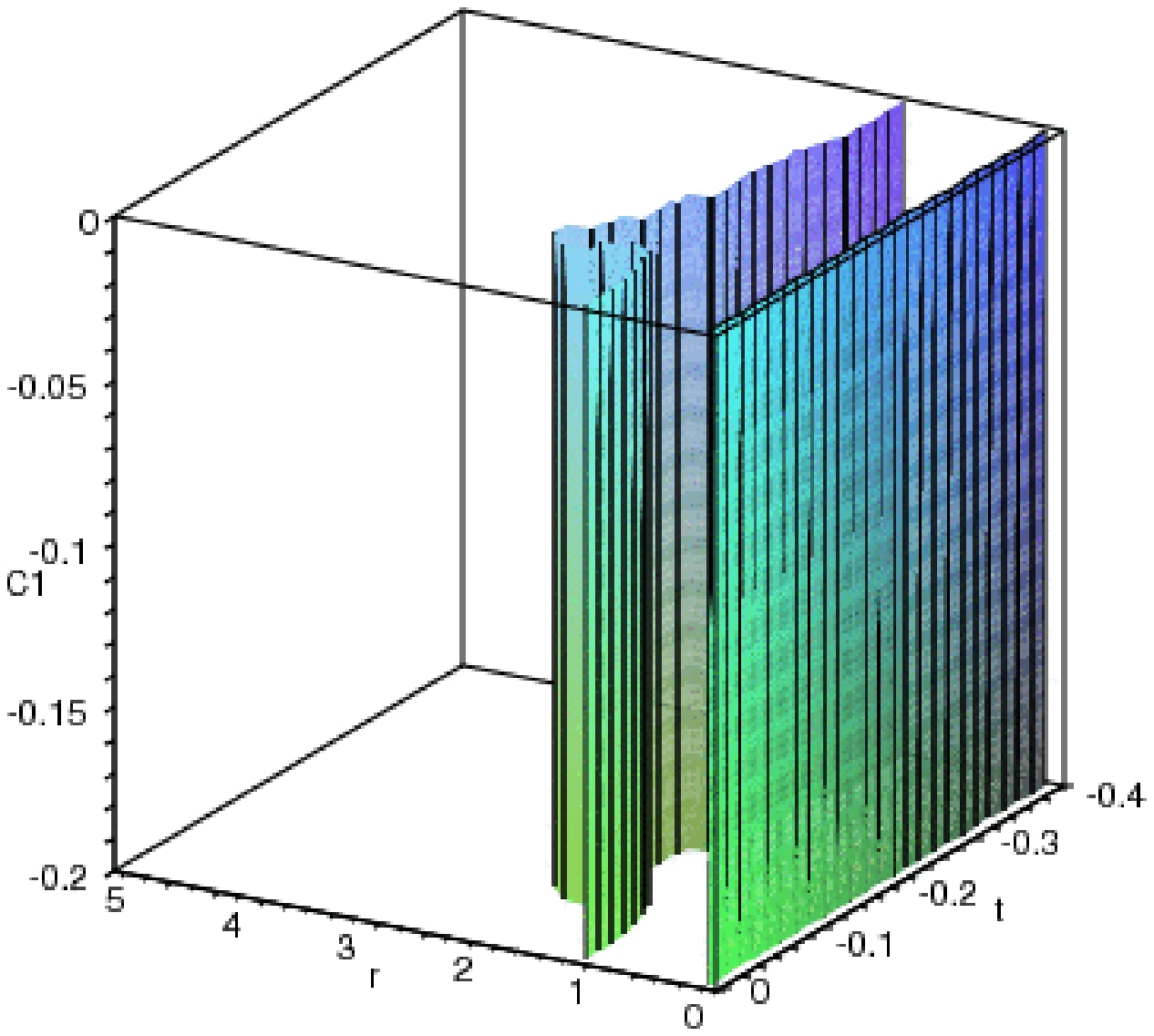,width=3.0truein,height=3.0truein}
\hskip .1in   \psfig{figure=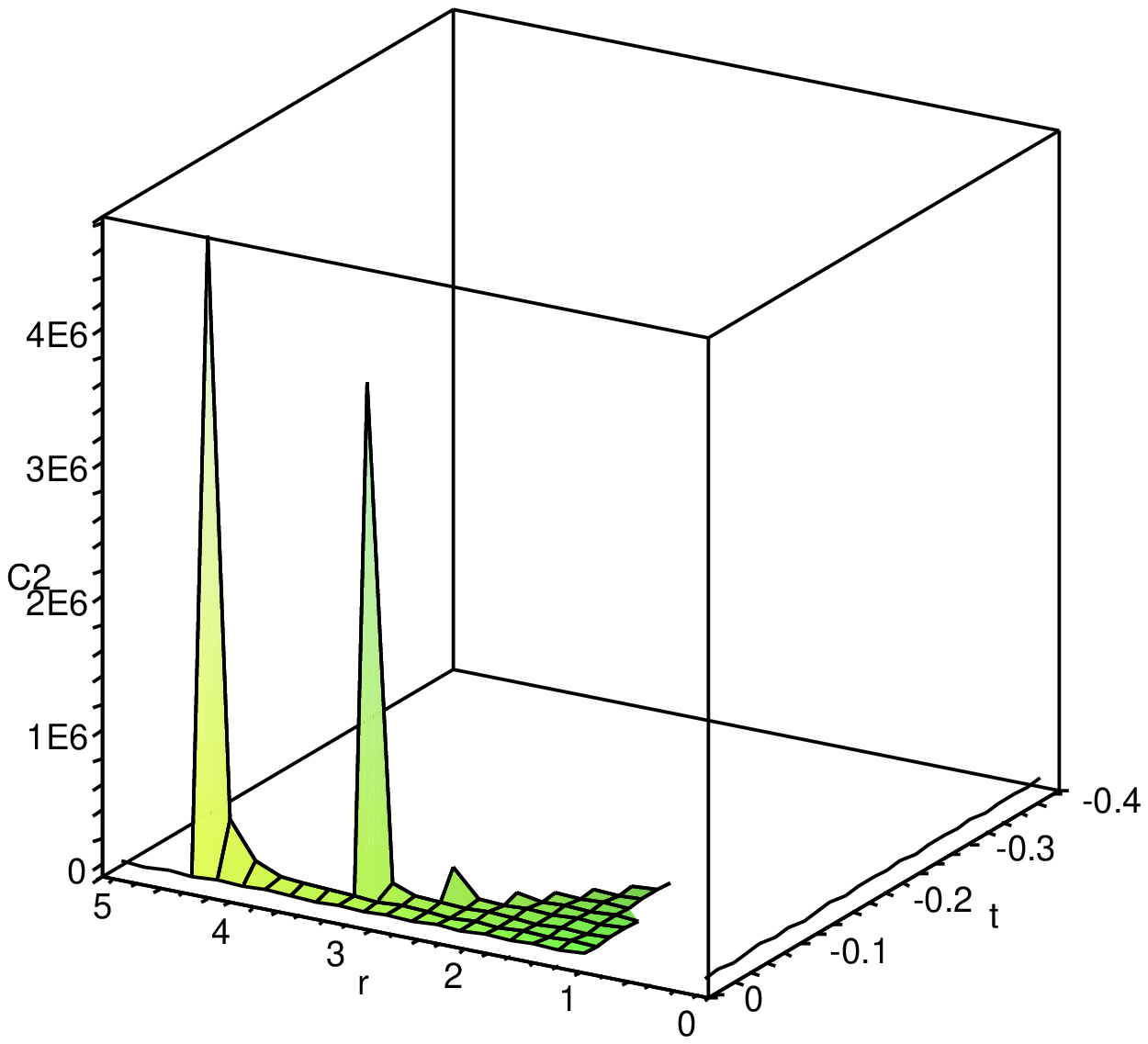,width=3.0truein,height=3.0truein}
\hskip .5in}
\centerline{  \psfig{figure=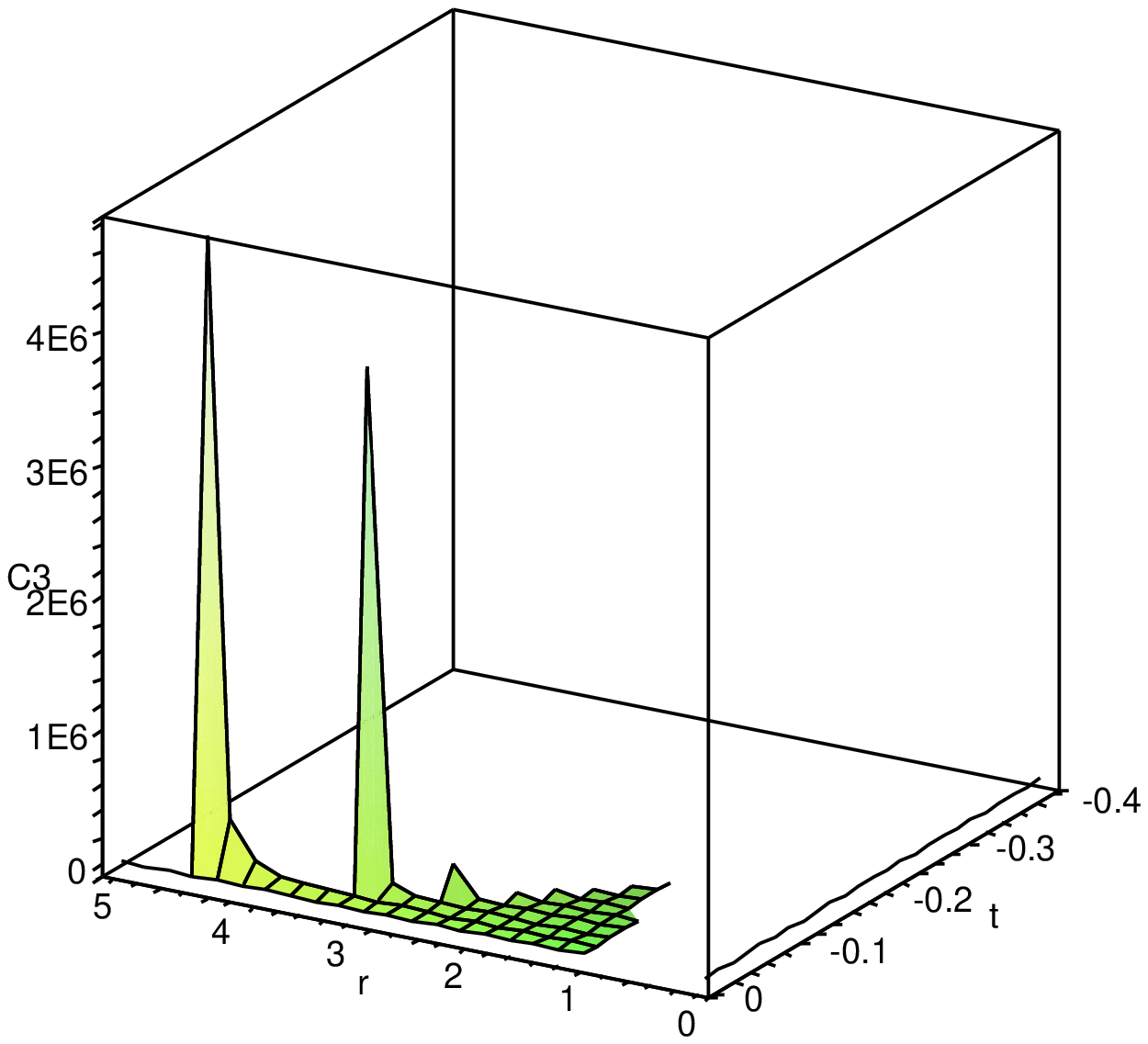,width=3.0truein,height=3.0truein}
\hskip .1in   \psfig{figure=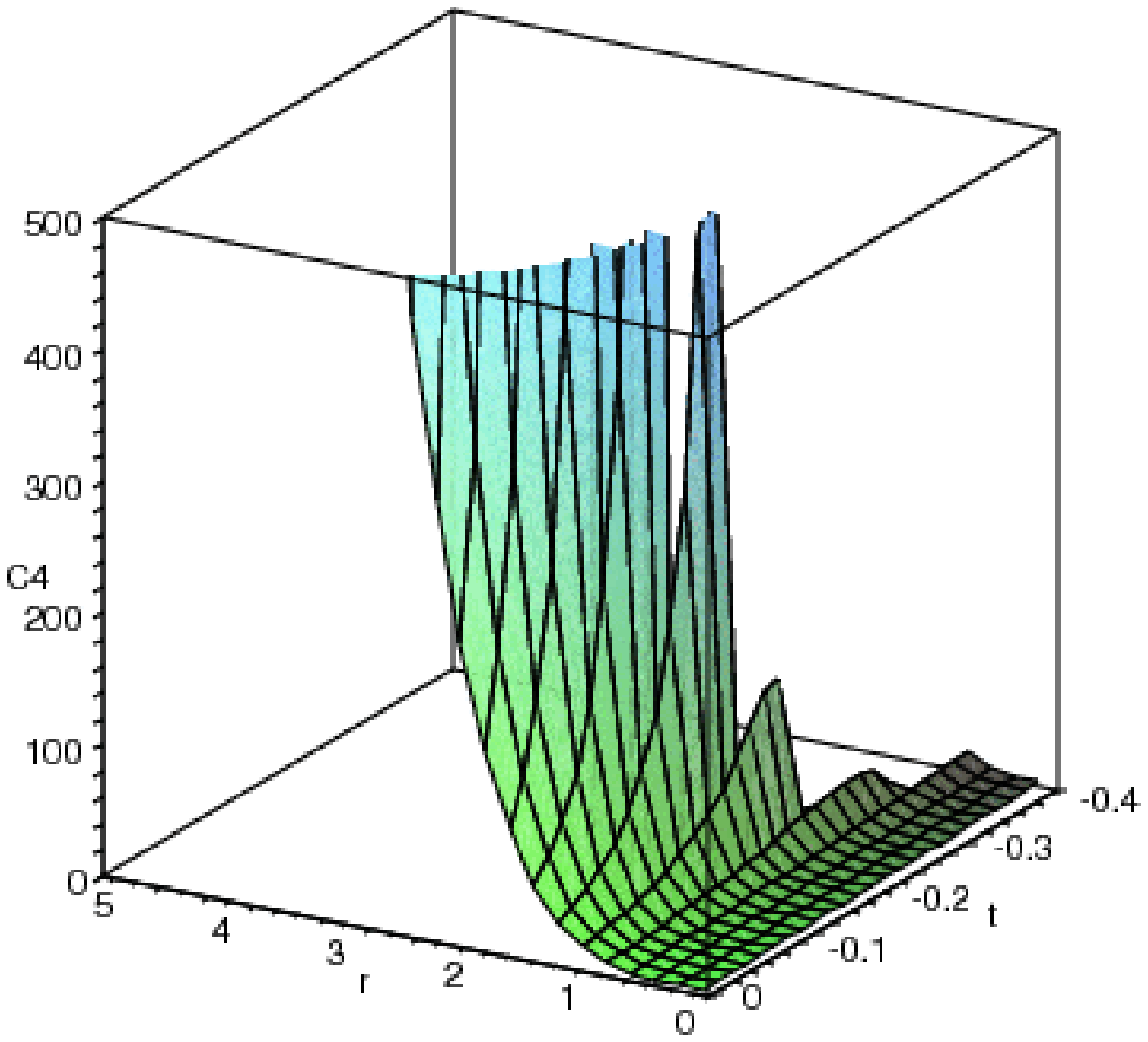,width=3.0truein,height=3.0truein}
\hskip .5in}
\centerline{  \psfig{figure=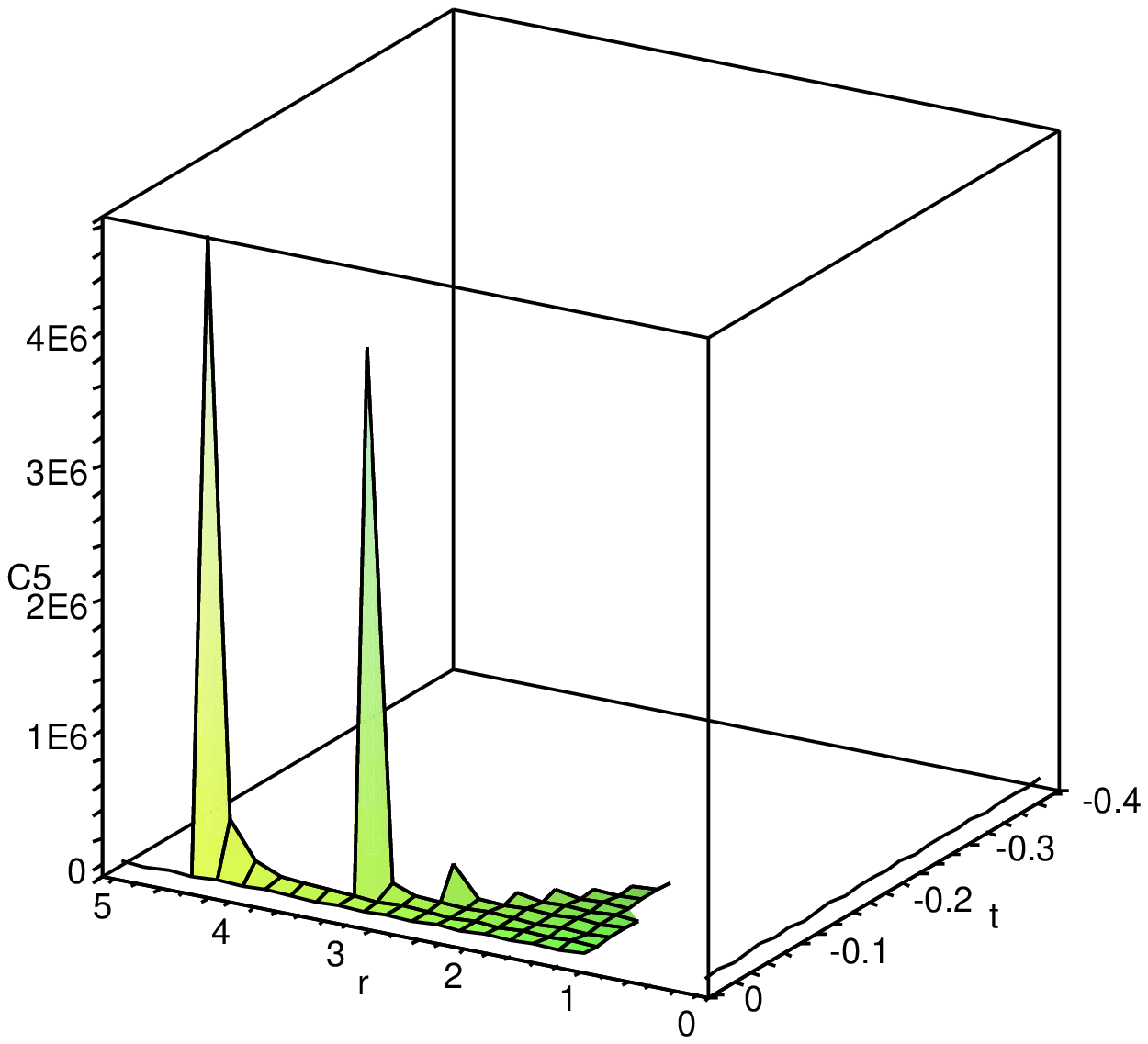,width=3.0truein,height=3.0truein}
\hskip .1in   \psfig{figure=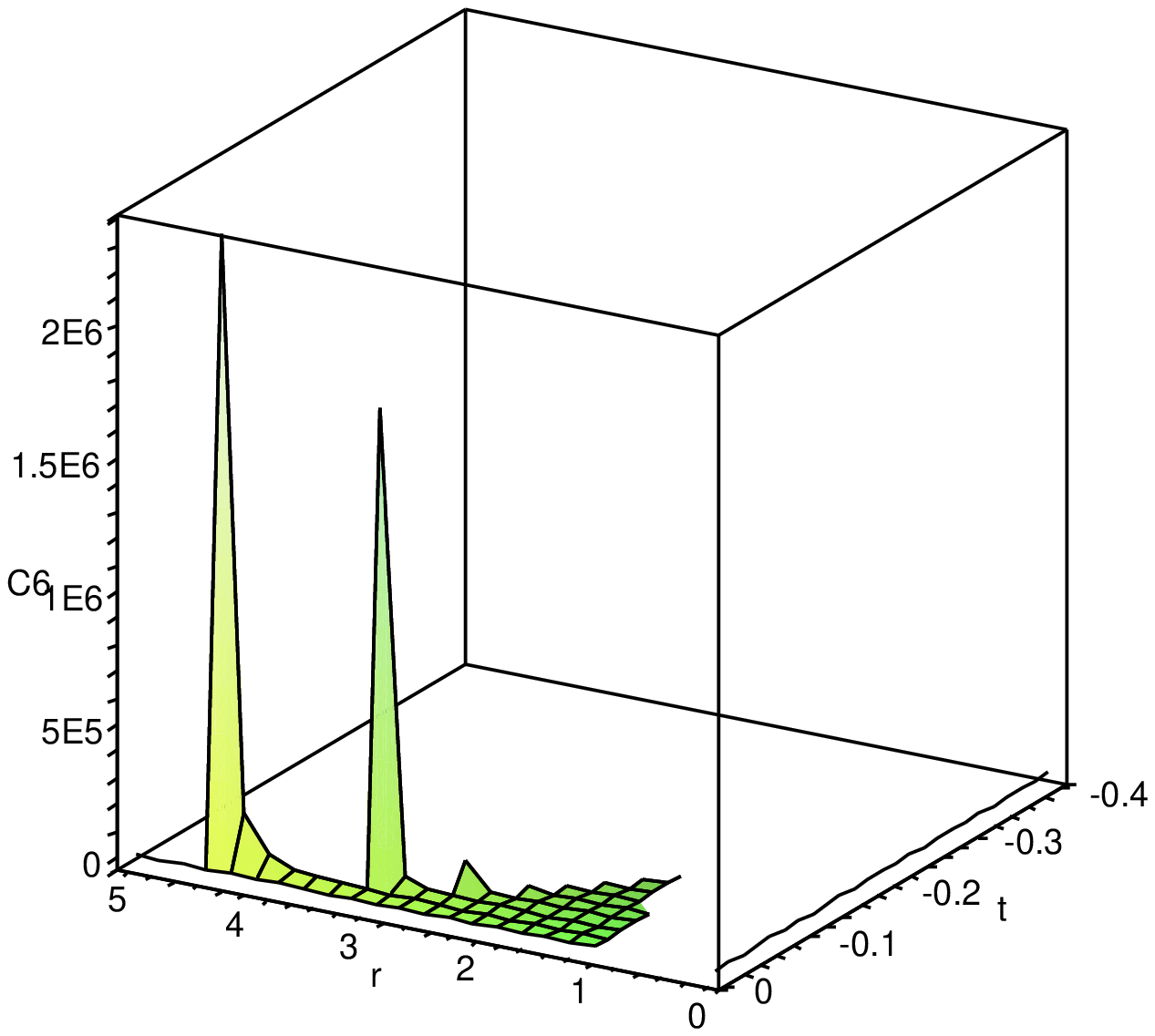,width=3.0truein,height=3.0truein}
	\hskip .5in}  \caption{The energy conditions for $\alpha=-2$ and $x_0=0$.}
\label{condalpham2}
\end{figure}

\begin{figure}
\vspace{.2in}
\centerline{  \psfig{figure=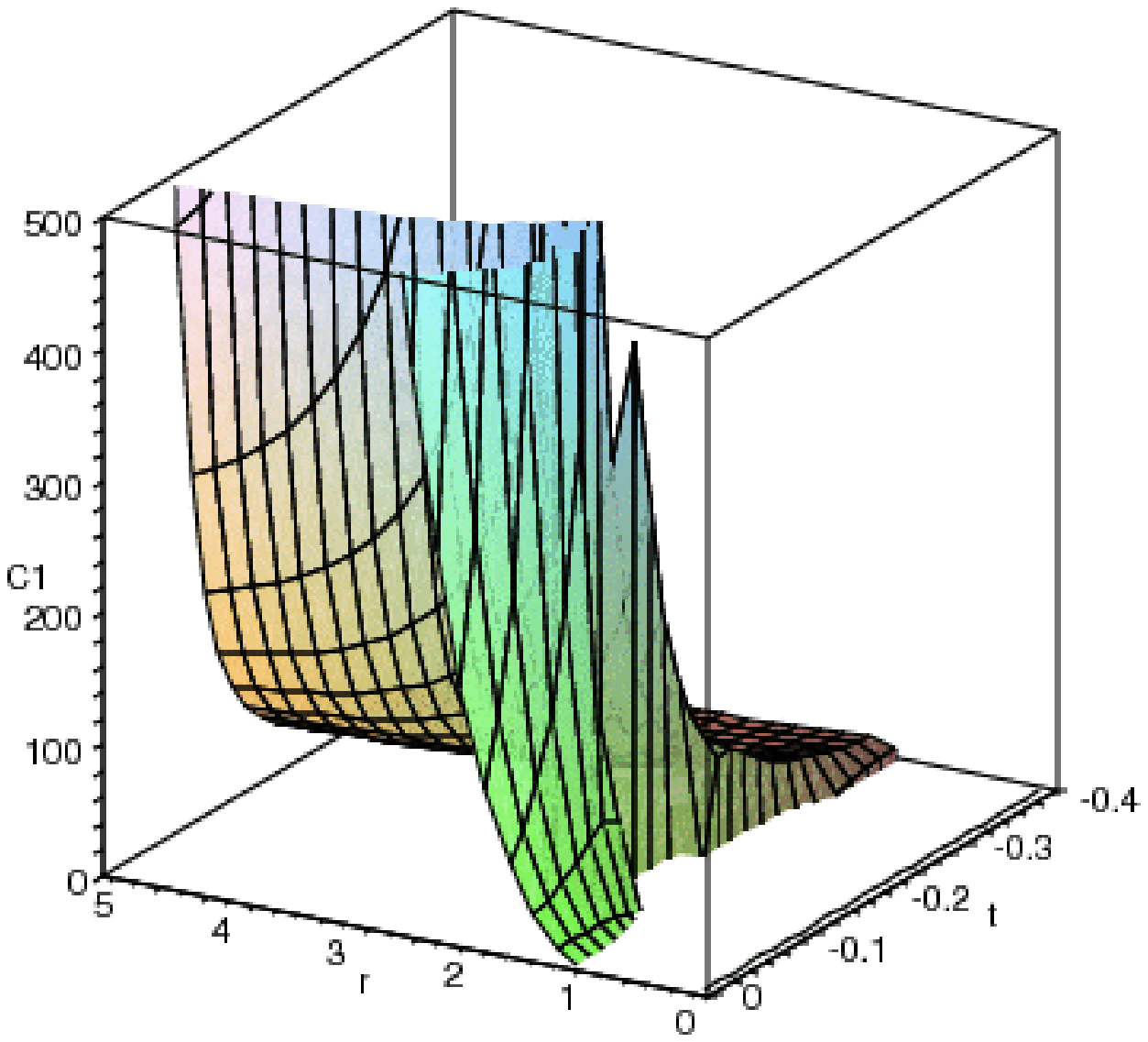,width=3.0truein,height=3.0truein}
\hskip .1in   \psfig{figure=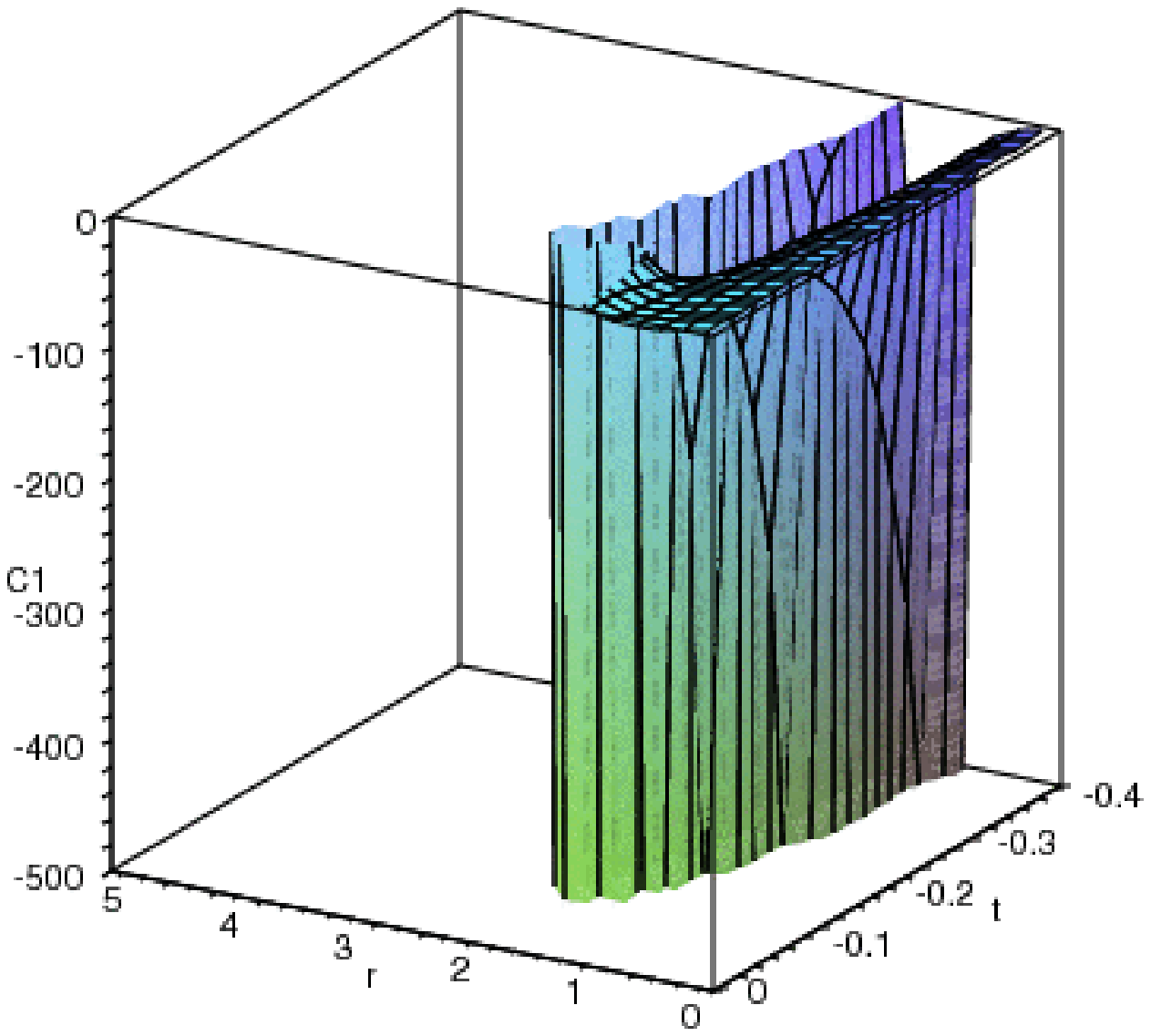,width=3.0truein,height=3.0truein}
\hskip .5in}  \caption{The positive and negative part of the energy condition 
$C1$ for $\alpha=-2$ and $x_0=0$.}
\label{condalpham2a}
\end{figure}

Again, we can see in Figure \ref{condalpham2} that all the energy conditions are 
satisfied except the Condition 1 ($C_1$) 
in the neighborhood
 of the curves where $C_1=0$ 
(see also Figure \ref{condalpham2a}).

\section{Conclusion}

In this work we have studied the evolution of an
anisotropic shear-free fluid with heat flux fluid and self-similarity of 
the second
kind, in order to build gravitational collapse model.
We have found a class of solution to the Einstein field equations by
assuming that the part of the tangential pressure which is explicitly time
dependent of the fluid is zero
and that the fluid moves along time-like geodesics.
The energy conditions, geometrical and physical properties
of the solutions were studied. We have also found that, depending on the 
parameter 
$\alpha$ and the geometrical radius, they may represent a 
naked singularity. 
Besides, the energy conditions are satisfied almost everywhere.

Comparing this collapsing system of imperfect fluid with heat flow with
the solution obtained by Chan, da Silva  \& Villas da Rocha \cite{CdSVR}, we can
see an important similarity, because in both we have self-similarity,
although corresponding to different kind, and shear-free configuration,
with naked singularity formation. The present results reinforce our
conclusions in the first paper, that is, although Joshi, Dadhich \&
Maartens \cite{JoshiD} concluded that the formation of naked singularities 
is due to
shear of the fluid, in such mode that sufficiently strong shearing effects
could delay the formation of apparent horizons, thereby exposing the strong
gravitational regions to the outside world and leading to naked
singularities, their study is based on a gravitational collapse of
spherically symmetric dust fluid. In another work more recent, Joshi,
Malafarina \& Saraykar \cite{JoshiM} showed that the causal structure of the
spacetime is in fact affected by the introducing of a small amount of
pressure. Our present results, again seem point out that the presence of
 anisotropic pressures can really modify the final structure in the
gravitational collapse. In addition, Giambo \& Magli, in a recent paper
\cite{Giambo}, studied the gravitational collapse of perfect fluids,  more
specifically for linear barotropic  fluids and, as in our case, a family of
geodesic fluid, and they concluded that the pressure plays a relevant role
on the causal structure of the collapsing model.

Moreover, the fluid here evolves to an exotic fluid near the formation of
the apparent horizon. Considering this kind of fluid can impose a
repulsive gravitational effect
(due to the phantom energy)
, we speculate that this can be the reason to
the inexistence of the formation of a horizon. 

Finally, it would be very interesting verify if the naked singularity could 
represent a critical solution for the gravitational collapse.

\section*{Acknowledgments}
                                                                                
The financial assistance from
FAPERJ/UERJ (MFAdaS and CFCB) is gratefully acknowledged. The
author (RC) acknowledges the financial support from FAPERJ (no.
E-26/171.754/2000, E-26/171.533/2002 and E-26/170.951/2006).
MFAdaS and RC also acknowledge the financial support from
Conselho Nacional de Desenvolvimento Cient\'{\i}fico e Tecnol\'ogico -
CNPq - Brazil.  The author (MFAdaS) also acknowledges the financial support
from Financiadora de Estudos e Projetos - FINEP - Brazil (Ref. 2399/03) 
, CNPq (Edital Universal 477268/2010-2) and FAPERJ (E-26/111.714/2010).
We would like also to thank Dr. A.Y. Miguelote for helpful discussions
at the beginning of this work.

\end{document}